\documentclass[11pt]{article}
\usepackage{amssymb,amsmath,amsfonts}
\usepackage{graphicx}
\usepackage{graphics}
\usepackage{eepic,epsfig}

\@addtoreset{equation}{section}

\makeatother

\begin{document}

\title{Fermionic vacuum polarization by a cylindrical boundary in the cosmic
string spacetime }
\author{E. R. Bezerra de Mello$^{1}$\thanks{%
E-mail: emello@fisica.ufpb.br},\, V. B. Bezerra$^{1}$\thanks{%
E-mail: valdir@fisica.ufpb.br}, \, A. A. Saharian$^{1,2}$\thanks{%
E-mail: saharian@ictp.it}, A. S. Tarloyan$^{2,3}$ \\
\\
\textit{$^{1}$Departamento de F\'{\i}sica, Universidade Federal da Para\'{\i}%
ba}\\
\textit{58.059-970, Caixa Postal 5.008, Jo\~{a}o Pessoa, PB, Brazil}\vspace{%
0.3cm}\\
\textit{$^2$Department of Physics, Yerevan State University,}\\
\textit{0025 Yerevan, Armenia}\vspace{0.3cm}\\
\textit{$^3$Yerevan Physics Institute, 0036 Yerevan, Armenia}}
\maketitle

\begin{abstract}
The vacuum expectation values of the energy--momentum tensor and the
fermionic condensate are analyzed for a massive spinor field obeying the MIT
bag boundary condition on a cylindrical shell in the cosmic string
spacetime. Both regions inside and outside the shell are considered. By
applying to the corresponding mode-sums a variant of the generalized
Abel--Plana formula, we explicitly extract the parts in the expectation
values corresponding to the cosmic string geometry without boundaries. In
this way the renormalization procedure is reduced to that for the
boundary-free cosmic string spacetime. The parts induced by the cylindrical
shell are presented in terms of integrals rapidly convergent for points away
from the boundary. The behavior of the vacuum densities is investigated in
various asymptotic regions of the parameters. In the limit of large values
of the planar angle deficit, the boundary-induced expectation values are
exponentially suppressed. As a special case, we discuss the fermionic vacuum
densities for the cylindrical shell on the background of the Minkowski
spacetime.
\end{abstract}

\bigskip


\bigskip

\section{Introduction}

It is well known that different types of topological objects may have been
formed in the early universe after Planck time by the vacuum phase
transition \cite{Vile85}. Depending on the topology of the vacuum manifold
these are domain walls, strings, monopoles and textures. Among them the
cosmic strings are of special interest. Although the recent observational
data on the cosmic microwave background radiation have ruled out cosmic
strings as the primary source for primordial density perturbations, they are
still candidates for the generation of a number of interesting physical
effects such as the generation of gravitational waves, gamma ray bursts and
high-energy cosmic rays (see, for instance, \cite{Damo00}). Recently, cosmic
strings attract a renewed interest partly because a variant of their
formation mechanism is proposed in the framework of brane inflation \cite%
{Sara02}.

In the simplest theoretical model describing the infinite straight cosmic
string the spacetime is locally flat except on the string where it has a
delta shaped Riemann curvature tensor. In quantum field theory the
corresponding non-trivial topology induces non-zero vacuum expectation
values for physical observables. Explicit calculations for the geometry of a
single cosmic string have been done for different fields \cite{Hell86}-\cite%
{Spin08}. Vacuum polarization effects by higher-dimensional composite
topological defects constituted by a cosmic string and global monopole are
investigated in Refs. \cite{Beze06Comp} for scalar and fermionic fields.
Another type of vacuum polarization arises when boundaries are present. The
imposed boundary conditions on quantum fields alter the zero-point
fluctuations spectrum and result in additional shifts in the vacuum
expectation values of physical quantities. This is the well-known Casimir
effect (for a review see \cite{Most97}). In Ref. \cite{Beze06sc}, we have
studied both types of sources for the polarization of the scalar vacuum,
namely, a cylindrical boundary and a cosmic string, assuming that the
boundary is coaxial with the string and that on this surface the scalar
field obeys Robin boundary condition. For a massive scalar field with an
arbitrary curvature coupling parameter we evaluated the Wightman function
and the vacuum expectation values of the field squared and the
energy-momentum tensor. The polarization of the electromagnetic vacuum by a
conducting cylindrical shell in the cosmic string spacetime is investigated
in \cite{Beze07El} (For a combination of topological and boundary induced
quantum effects in the gravitational field of a global monopole see Refs.
\cite{Saha03Mon,Saha04Mon,Beze06Mon}.)

Continuing in this line of investigation, in the present paper we analyze
the polarization of the fermionic vacuum by a cylindrical shell coaxial with
the cosmic string on which the field obeys MIT bag boundary condition. We
evaluate the fermionic condensate and vacuum expectation values of the
energy-momentum tensor in both interior and exterior regions of the shell.
The renormalized vacuum expectation value of the energy-momentum tensor for
a fermionic field in the geometry of a cosmic string without boundaries is
investigated in \cite{Frol87,Dowk87b,Line95,More95,Beze06}. In addition to
describing the physical structure of the quantum field at a given point, the
energy-momentum tensor acts as a source of gravity in the Einstein equations
and plays an important role in modelling a self-consistent dynamics
involving the gravitational field. In the problem under consideration all
calculations can be performed in a closed form and it constitutes an example
in which the topological and boundary-induced polarizations of the vacuum
can be separated in different contributions.

From the point of view of the physics in the region outside the string, the
geometry considered in the present paper can be viewed as a simplified model
for the non-trivial core. This model presents a framework in which the
influence of the finite core effects on physical processes in the vicinity
of the cosmic string can be investigated. In particular, it enables to
specify conditions under which the idealized model with the core of zero
thickness can be used. The corresponding results may shed light \ upon
features of finite core effects in more realistic models, including those
used for string-like defects in crystals and superfluid helium. In addition,
the problem considered here is of interest as an example with combined
topological and boundary induced quantum effects in which the vacuum
characteristics such as energy density and stresses can be found in closed
analytic form. From the results of the present paper, as a special case, we
obtain the fermionic Casimir densities for a cylindrical shell with MIT
boundary conditions in the Minkowski background (for the combined effects of
a magnetic fluxon and MIT boundary conditions on the vacuum energy of a
Dirac field see Refs. \cite{Lese98}). Note that, in addition to traditional
problems of quantum field theory under the presence of material boundaries,
the Casimir effect for cylindrical geometries can also be important to the
flux tube models of confinement \cite{Fish87,Barb90} and for determining the
structure of the vacuum state in interacting field theories \cite{Ambj83}.

We have organized the paper as follows. In the next section, the
eigenspinors for the region inside the cylindrical boundary are constructed
and the eigenvalues of the corresponding quantum numbers are specified.
These eigenspinors are the basis for the anlysis of the Casimir densities in
the following sections. In section \ref{sec:FermC}, by using a variant of
the generalized Abel-Plana formula, we extract from the mode-sum of the
fermionic condensate the part corresponding to the geometry of a cosmic
string without the shell. The part induced by the shell is investigated in
various asymptotic regions for the parameters. Section \ref{sec:EMT} is
devoted to the investigation of the boundary induced parts in the vacuum
expectation value of the energy-momentum tensor inside the cylindrical
shell. The vacuum densities in the region outside the cylindrical shell are
discussed in section \ref{sec:Exterior}. The main results of the paper are
summarized in section \ref{sec:Conc}. In appendix \ref{sec:Appendix} we give
an alternative representation for the fermionic condensate and the
expectation value of the energy-momentum tensor for a massive fermionic
field in the geometry of a cosmic string without boundaries.

\section{Eigenspinors inside a cylindrical shell}

\label{sec:EigSpinor}

We consider the background spacetime corresponding to an infinitely long
straight cosmic string with the conical line element%
\begin{equation}
ds^{2}=dt^{2}-dr^{2}-r^{2}d\phi ^{2}-dz{}^{2},  \label{ds21}
\end{equation}%
where $0\leqslant \phi \leqslant \phi _{0}\leqslant 2\pi $ and the spatial
points $(r,\phi ,z)$ and $(r,\phi +\phi _{0},z)$ are to be identified. The
planar angle deficit is related to the mass per unit length of the string, $%
\mu _{0}$, by $2\pi -\phi _{0}=8\pi G\mu _{0}$, where $G$ is the Newton
gravitational constant. In the discussion below, in addition to the
parameter $\phi _{0}$, we will use the combination%
\begin{equation}
q=2\pi /\phi _{0}.  \label{q}
\end{equation}

The dynamics of a massive fermionic field is governed by the Dirac equation
\begin{equation}
i\gamma ^{\mu }\nabla _{\mu }\psi -m\psi =0\ ,  \label{Direq}
\end{equation}%
with the covariant derivative operator defined as
\begin{equation}
\nabla _{\mu }=\partial _{\mu }+\Gamma _{\mu }\ .  \label{Covder}
\end{equation}%
Here $\gamma ^{\mu }=e_{(a)}^{\mu }\gamma ^{(a)}$ are the Dirac matrices in
curved spacetime and $\Gamma _{\mu }$ is the spin connection given in terms
of the flat space Dirac matrices $\gamma ^{(a)}$ by the relation
\begin{equation}
\Gamma _{\mu }=\frac{1}{4}\gamma ^{(a)}\gamma ^{(b)}e_{(a)}^{\nu }e_{(b)\nu
;\mu }\ .  \label{Gammamu}
\end{equation}%
Note that in this formula $;$ means the standard covariant derivative for
vector fields. In the relations above $e_{(a)}^{\mu }$ is the tetrad basis
satisfying $e_{(a)}^{\mu }e_{(b)}^{\nu }\eta ^{ab}=g^{\mu \nu }$, where $%
\eta ^{ab}$ is the Minkowski spacetime metric tensor.

In this paper we are interested in the change of the vacuum expectation
values (VEVs) of the fermionic condensate and the energy-momentum tensor for
a fermionic field, induced by a cylindrical shell coaxial \ with the string
on which the field obeys MIT bag boundary condition:
\begin{equation}
\left( 1+i\gamma ^{\mu }n_{\mu }\right) \psi =0\ ,\quad r=a,  \label{BagCond}
\end{equation}%
where $a$ is the cylinder radius and $n_{\mu }=(0,1,0,0)$ is the
outward-pointing normal to the boundary. For the evaluation of the VEVs we
will use the direct mode summation procedure. In this approach we need to
have the complete set of the eigenfunctions satisfying boundary condition (%
\ref{BagCond}).

In order to find these eigenfunctions, we will use the standard
representation of the flat space Dirac matrices:%
\begin{equation}
\gamma ^{(0)}=\left(
\begin{array}{cc}
1 & 0 \\
0 & -1%
\end{array}%
\right) ,\;\gamma ^{(a)}=\left(
\begin{array}{cc}
0 & \sigma _{a} \\
-\sigma _{a} & 0%
\end{array}%
\right) ,\;a=1,2,3,  \label{gam0l}
\end{equation}%
with $\sigma _{1},\sigma _{2},\sigma _{3}$ being the Pauli matrices. We take
the tetrad fields in the form used before in \cite{Sous89} (see also \cite%
{Beze06,Skar94}):%
\begin{equation}
e_{(a)}^{\mu }=\left(
\begin{array}{cccc}
1 & 0 & 0 & 0 \\
0 & \cos (q\phi ) & -\sin (q\phi )/r & 0 \\
0 & \sin (q\phi ) & \cos (q\phi )/r & 0 \\
0 & 0 & 0 & 1%
\end{array}%
\right) ,  \label{emua}
\end{equation}%
where the index $a$ identifies the rows of the matrix. With this choice, the
gamma matrices are given by%
\begin{equation}
\gamma ^{0}=\gamma ^{(0)},\;\gamma ^{l}=\left(
\begin{array}{cc}
0 & \beta ^{l} \\
-\beta ^{l} & 0%
\end{array}%
\right) ,\;l=1,2,3,  \label{gamcurved}
\end{equation}%
where we have introduced the $2\times 2$ matrices%
\begin{equation}
\beta ^{1}=\left(
\begin{array}{cc}
0 & e^{-iq\phi } \\
e^{iq\phi } & 0%
\end{array}%
\right) ,\;\beta ^{2}=-\frac{i}{r}\left(
\begin{array}{cc}
0 & e^{-iq\phi } \\
-e^{iq\phi } & 0%
\end{array}%
\right) ,\;\;\beta ^{3}=\left(
\begin{array}{cc}
1 & 0 \\
0 & -1%
\end{array}%
\right) .  \label{betl}
\end{equation}%
For the spin connection and the combination appearing in the Dirac equation
we find%
\begin{equation}
\Gamma _{\mu }=\frac{1-q}{2}\gamma ^{(1)}\gamma ^{(2)}\delta _{\mu
}^{2},\;\gamma ^{\mu }\Gamma _{\mu }=\frac{1-q}{2r}\gamma ^{1},
\label{gammu}
\end{equation}%
and the Dirac equation takes the form%
\begin{equation}
\left( \gamma ^{\mu }\partial _{\mu }+\frac{1-q}{2r}\gamma ^{1}+im\right)
\psi =0.  \label{Direq1}
\end{equation}

For positive frequency solutions, assuming the time-dependence of the
eigenfunctions in the form $e^{-i\omega t}$ and decomposing the bispinor $%
\psi $ into the upper and lower components, denoted by $\varphi $ and $\chi $%
, respectively, we find the equations%
\begin{eqnarray}
\left( \beta ^{l}\partial _{l}\mathbf{+}\frac{1-q}{2r}\beta ^{1}\right)
\varphi -i\left( \omega +m\right) \chi  &=&0,  \notag \\
\left( \beta ^{l}\partial _{l}\mathbf{+}\frac{1-q}{2r}\beta ^{1}\right) \chi
-i\left( \omega -m\right) \varphi  &=&0.  \label{phixieq}
\end{eqnarray}%
Substituting the function $\chi $ from the first equation into the second
one, we obtain the second order differential equation for the function $%
\varphi $:%
\begin{equation}
\bigg[-g^{nl}\partial _{n}\partial _{l}\mathbf{+}\frac{1}{r}\partial _{1}+%
\frac{q-1}{r}\beta ^{1}\beta ^{2}\partial _{2}\mathbf{-}\frac{(q-1)^{2}}{%
4r^{2}}+\omega ^{2}-m^{2}\bigg]\varphi =0,  \label{phieq}
\end{equation}%
where $n,l=1,2,3$. The same equation is obtained for the function $\chi $.

Because the above equation is in diagonal matrix form, we decompose the
spinor $\varphi $ into the upper, $\varphi _{1}$, and lower, $\varphi _{2}$,
components. Taking the eigenfunctions corresponding to these components in
the form $\varphi _{l}=R_{l}(r)e^{i(qn_{l}\phi +kz-\omega t)}$, $l=1,2$, we
can see that the solutions of the equations for the radial functions,
regular on the string, are expressed in terms of the Bessel function of the
first kind: $R_{l}(r)=C_{l}J_{\beta _{l}}(\lambda r)$, where the order is
defined by the relations%
\begin{equation}
\beta _{1}=|qn_{1}+(q-1)/2|,\;\beta _{2}=|qn_{2}-(q-1)/2|,  \label{bet12}
\end{equation}%
with $n_{1,2}=0,\pm 1,\pm 2,\ldots $, and%
\begin{equation}
\lambda =\sqrt{\omega ^{2}-k^{2}-m^{2}}.  \label{lambda}
\end{equation}%
Hence, the components of the upper spinor are given by the formula%
\begin{equation}
\varphi _{l}=C_{l}J_{\beta _{l}}(\lambda r)\exp \left[ i\left( qn_{l}\phi
+kz-\omega t\right) \right] ,\;l=1,2.  \label{phil}
\end{equation}%
Having the upper spinor, we can find the components $\chi _{l}$ of the lower
one by using the first equation in (\ref{phixieq}). From this equation we
find the following relations%
\begin{equation}
n_{2}=n_{1}+1,\;\beta _{2}=\beta _{1}+\epsilon _{n_{1}},\;\epsilon
_{n_{1}}\equiv \mathrm{sgn}(n_{1}),  \label{n21rel}
\end{equation}%
and%
\begin{equation}
\chi _{l}=B_{l}J_{\beta _{l}}(\lambda r)\exp \left[ i\left( qn_{l}\phi
+kz-\omega t\right) \right] ,\;l=1,2,  \label{xil}
\end{equation}%
with the coefficients%
\begin{equation*}
B_{1}=\frac{kC_{1}-i\epsilon _{n_{1}}\lambda C_{2}}{\omega +m},\;B_{2}=-%
\frac{kC_{2}-i\epsilon _{n_{1}}\lambda C_{1}}{\omega +m}.
\end{equation*}%
We can see that the bispinor with the components defined by relations (\ref%
{phil}) and (\ref{xil}) is an eigenfunction of the projection of the total
momentum along the cosmic string:%
\begin{equation}
\widehat{J}_{3}\psi =\left( -i\partial _{\phi }+i\frac{q}{2}\gamma
^{(1)}\gamma ^{(2)}\right) \psi =qj\psi ,  \label{J3}
\end{equation}%
where%
\begin{equation}
j=n_{1}+1/2,\;j=\pm 1/2,\pm 3/2,\ldots .  \label{j}
\end{equation}

For the further specification of the eigenfunctions we can impose an
additional condition relating the constants $C_{1}$ and $C_{2}$. As such a
condition, following \cite{Bord96}, we will require the following relations
between the upper and lower components:%
\begin{equation}
\chi _{1}=\kappa \varphi _{1},\;\chi _{2}=-\frac{\varphi _{2}}{\kappa }.
\label{xiphisim}
\end{equation}%
From the expressions for the spinor components we find the eigenvalues of
the parameter $\kappa $,%
\begin{equation}
\kappa =\kappa _{s}\equiv \frac{\omega +s\sqrt{\omega ^{2}-k^{2}}}{k}%
,\;s=\pm 1,  \label{kappas}
\end{equation}%
and the relation%
\begin{equation}
C_{2}=i\epsilon _{n_{1}}\frac{\kappa _{s}}{\lambda }(m+s\sqrt{\omega
^{2}-k^{2}})C_{1},  \label{C21rel}
\end{equation}%
for the coefficients in (\ref{phil}).

Hence, the positive frequency solutions to the Dirac equation, specified by
the set of quantum numbers $\sigma =(\lambda ,j,k,s)$, has the form%
\begin{equation}
\psi _{\sigma }^{(+)}=C_{\sigma }^{(+)}\left(
\begin{array}{c}
J_{\beta _{1}}(\lambda r) \\
i\epsilon _{j}\kappa _{s}b_{s}^{(+)}J_{\beta _{2}}(\lambda r)e^{iq\phi } \\
\kappa _{s}J_{\beta _{1}}(\lambda r) \\
-i\epsilon _{j}b_{s}^{(+)}J_{\beta _{2}}(\lambda r)e^{iq\phi }%
\end{array}%
\right) \exp \left[ i\left( q(j-1/2)\phi +kz-\omega t\right) \right] ,
\label{psi+}
\end{equation}%
where the orders of the Bessel functions are defined in terms of $j$ as%
\begin{eqnarray}
\beta _{1} &=&|qj-1/2|=q|j|-\epsilon _{j}/2,  \notag \\
\beta _{2} &=&|qj+1/2|=q|j|+\epsilon _{j}/2.  \label{bet12a}
\end{eqnarray}%
In (\ref{psi+}) and in the consideration below we use the notations%
\begin{equation}
b_{s}^{(\pm )}=\frac{\pm m+s\sqrt{\lambda ^{2}+m^{2}}}{\lambda }.
\label{bsplmin}
\end{equation}%
Note that one has the relation $b_{s}^{(\mp )}=1/b_{s}^{(\pm )}$.

The eigenvalues of the radial quantum number $\lambda $ are determined from
the boundary condition (\ref{BagCond}) imposed on the eigenspinor (\ref{psi+}%
). For fixed values of $j$ and $s$, this leads to the single equation%
\begin{equation}
J_{\beta _{1}}(\lambda a)+\epsilon _{j}b_{s}^{(+)}J_{\beta _{2}}(\lambda
a)=0.  \label{lambvalue}
\end{equation}%
Using the recurrence relations for the Bessel functions, this equation may
also be written in the form%
\begin{equation}
\tilde{J}_{\beta _{1}}(\lambda a)=0.  \label{lambvalue1}
\end{equation}%
Here and in what follows we use the notation%
\begin{eqnarray}
\tilde{J}_{\beta _{1}}(x) &=&xJ_{\beta _{1}}^{\prime }(x)+(\mu -s\sqrt{%
x^{2}+\mu ^{2}}-\epsilon _{j}\beta _{1})J_{\beta _{1}}(x)  \notag \\
&=&-x\epsilon _{j}[J_{\beta _{2}}(x)+\epsilon _{j}b_{s}^{(-)}J_{\beta
_{1}}(\lambda a)],  \label{ftilde}
\end{eqnarray}%
with $\mu =ma$. We will denote the solutions of the equation (\ref%
{lambvalue1}) by $\lambda a=\lambda _{\beta _{1},l}$, $l=1,2,\ldots $,
assuming that they are arranged in ascending order. Now the set of quantum
numbers is specified by $\sigma =(l,j,k,s)$.

The coefficient $C_{\sigma }^{(+)}$ in (\ref{psi+}) is determined from the
normalization condition
\begin{equation}
\int d^{3}x\sqrt{\gamma }\psi _{\sigma }^{(+)+}\psi _{\sigma ^{\prime
}}^{(+)}=\delta _{\sigma \sigma ^{\prime }}\ ,  \label{normcond}
\end{equation}%
where $\gamma $ is the determinant of the spatial metric and the integration
goes over the region inside the cylindrical shell. The delta symbol on the
right-hand side of Eq. (\ref{normcond}) is understood as the Dirac delta
function for continuous quantum numbers ($k$) and the Kronecker delta for
discrete ones ($l,j,s$). Substituting the eigenspinors (\ref{psi+}) into Eq.
(\ref{normcond}) and using the value of the standard integral involving the
square of the Bessel function \cite{Prud86}, we find%
\begin{equation}
(C_{\sigma }^{(+)})^{-2}=2\pi \phi _{0}a^{2}J_{\beta _{1}}^{2}(x)\frac{%
\kappa _{s}^{2}+1}{x^{2}}[2(x^{2}+\mu ^{2})+s\left( 2\beta _{1}\epsilon
_{j}+1\right) \sqrt{x^{2}+\mu ^{2}}+\mu ],  \label{Csig+}
\end{equation}%
with the notation $x=\lambda a$.

The negative frequency eigenspinors are found in a way similar to that used
above for the positive frequency ones and have the form%
\begin{equation}
\psi _{\sigma }^{(-)}=C_{\sigma }^{(-)}\left(
\begin{array}{c}
J_{\beta _{2}}(\lambda r) \\
i\epsilon _{j}\kappa _{s}b_{s}^{(-)}J_{\beta _{1}}(\lambda r)e^{iq\phi } \\
\kappa _{s}J_{\beta _{2}}(\lambda r) \\
-i\epsilon _{j}b_{s}^{(-)}J_{\beta _{1}}(\lambda r)e^{iq\phi }%
\end{array}%
\right) \exp \left[ -i\left( q(j+1/2)\phi +kz-\omega t\right) \right] ,
\label{psi-}
\end{equation}%
with the same notations as in (\ref{psi+}). The boundary condition imposed
on this eigenspinor leads to the same equation (\ref{lambvalue}) for the
eigenvalues of $\lambda $. The coefficient $C_{\sigma }^{(-)}$ is found from
the orthonormalization condition which is similar to (\ref{normcond}) and
has the form $C_{\sigma }^{(-)}=C_{\sigma }^{(+)}/b_{s}^{(-)}$.

\section{Fermionic condensate}

\label{sec:FermC}

Fermionic condensate is evaluated by using the mode-sum formula
\begin{equation}
\langle 0|\bar{\psi}\psi |0\rangle =\sum_{\sigma }\bar{\psi}_{\sigma
}^{(-)}\psi _{\sigma }^{(-)},  \label{FCmodesum}
\end{equation}%
where $\bar{\psi}_{\sigma }^{(-)}=\psi _{\sigma }^{(-)+}$ is the Dirac
adjoint and%
\begin{equation}
\sum_{\sigma }=\sum_{j=\pm 1/2,\pm 3/2,\cdots }\int_{-\infty }^{+\infty
}dk\,\sum_{s=\pm 1}\sum_{l=1}^{\infty }.  \label{Sumsig}
\end{equation}%
Substituting the eigenspinor (\ref{psi-}) into (\ref{FCmodesum}), we find
\begin{equation}
\langle 0|\bar{\psi}\psi |0\rangle =\sum_{\sigma }(\kappa
_{s}^{2}-1)C_{\sigma }^{(-)2}[b_{s}^{(-)2}J_{\beta }^{2}(\lambda r)-J_{\beta
+\epsilon _{j}}^{2}(\lambda r)]_{\lambda =\lambda _{\beta ,l}/a},
\label{FC1}
\end{equation}%
where $\kappa _{s}$ is defined by expression (\ref{kappas}) and
\begin{equation}
\beta =\beta _{1}=q|j|-\epsilon _{j}/2.  \label{beta}
\end{equation}%
The fermionic condensate given by formula (\ref{FC1}) is divergent and some
regularization procedure is necessary. We will assume that a cutoff function
is introduced in formula (\ref{FC1}) without explicitly writing it.

As the explicit form for $\lambda _{\beta ,l}$ is not known, formula (\ref%
{FC1}) is not convenient for the direct evaluation of the condensate. In
addition, the separate terms in the mode-sum are highly oscillatory for
large values of the quantum numbers. A convenient form can be obtained by
applying to the series over $l$ the summation formula, previously derived in
Ref. \cite{Saha04Mon} (see, also, \cite{Saha07}). In \cite{Saha04Mon} by
using the generalized Abel-Plana formula, it has been shown that for a
function $h(z)$ analytic in the half-plane $\mathrm{Re\,}z>0$ and satisfying
the condition
\begin{equation}
|h(z)|<\varepsilon (x)e^{c|y|}\ ,\quad z=x+iy,\quad |z|\rightarrow \infty \ ,
\label{condf}
\end{equation}%
with $c<2$ and $\varepsilon (x)\rightarrow 0$ for $x\rightarrow \infty $,
the following formula takes place
\begin{eqnarray}
&&{}\sum_{l=1}^{\infty }T_{\beta }(\lambda _{\beta ,l})h(\lambda _{\beta
,l})=\int_{0}^{\infty }h(x)dx+\frac{\pi }{2}\underset{z=0}{\mathrm{Res}}%
\bigg[h(z)\frac{\tilde{Y}_{\beta }(z)}{\tilde{J}_{\beta }(z)}\bigg]  \notag
\\
&&{}-\frac{1}{\pi }\int_{0}^{\infty }dx\,\bigg[e^{-\beta \pi i}h(xe^{\pi
i/2})\frac{K_{\beta }^{(+)}(x)}{I_{\beta }^{(+)}(x)}+e^{\beta \pi
i}h(xe^{-\pi i/2})\frac{K_{\beta }^{(-)}(x)}{I_{\beta }^{(-)}(x)}\bigg]\ ,
\label{SumLamb}
\end{eqnarray}%
where $I_{\beta }(x)$, $K_{\beta }(x)$ are the modified Bessel functions,
and $T_{\beta }(z)$ is defined by the relation
\begin{equation}
zT_{\beta }^{-1}(z)=J_{\beta }^{2}(z)\bigg[z^{2}+(\mu -\epsilon _{j}\beta
)(\mu ^{2}-s\sqrt{z^{2}+\mu ^{2}})+\frac{sz^{2}}{2\sqrt{z^{2}+\mu ^{2}}}%
\bigg]\ .  \label{r1}
\end{equation}%
In formula (\ref{SumLamb})\bigskip\ we used the notations
\begin{equation}
F^{(\pm )}(z)=\left\{
\begin{array}{cc}
zF^{\prime }(z)+(\mu -s\sqrt{\mu ^{2}-z^{2}}-\epsilon _{j}\beta )F(z)\
,\quad  & |z|<\mu , \\
zF^{\prime }(z)+(\mu \mp si\sqrt{z^{2}-\mu ^{2}}-\epsilon _{j}\beta )F(z)\
,\quad  & |z|>\mu ,%
\end{array}%
\right.   \label{Fbarpm}
\end{equation}%
for a given function $F(z)$.

By taking into account the relation%
\begin{equation}
C_{\sigma }^{(-)2}=\frac{k^{2}x}{8\pi \phi _{0}a\omega }\frac{\sqrt{%
x^{2}+\mu ^{2}}+s\mu }{a\omega +s\sqrt{x^{2}+\mu ^{2}}}\frac{T_{\beta }(x)}{%
\sqrt{x^{2}+\mu ^{2}}},  \label{CTrel}
\end{equation}%
we can write the mode-sum for the fermionic condensate in the form
\begin{equation}
\langle 0|\bar{\psi}\psi |0\rangle =-\frac{1}{4\pi \phi _{0}a^{2}}%
\sum_{\sigma }\frac{xT_{\beta }(x)}{a\omega }f_{\beta }(x,xr/a)|_{x=\lambda
_{\beta ,l}},  \label{FC2}
\end{equation}%
with the notation%
\begin{equation}
f_{\beta }(x,y)=(\mu -s\sqrt{x^{2}+\mu ^{2}})J_{\beta }^{2}(y)+(\mu +s\sqrt{%
x^{2}+\mu ^{2}})J_{\beta +\epsilon _{j}}^{2}(y).  \label{fxy}
\end{equation}%
Note that in (\ref{FC2})%
\begin{equation}
a\omega =\sqrt{x^{2}+k^{2}a^{2}+\mu ^{2}},  \label{aom}
\end{equation}%
and we have the property%
\begin{equation}
e^{-\beta \pi i}f_{\beta }(xe^{\pi i/2},ye^{\pi i/2})=e^{\beta \pi
i}f_{\beta }(xe^{-\pi i/2},ye^{-\pi i/2}),  \label{fprop}
\end{equation}%
for $x<\mu $. Now we apply to the sum over $l$ in (\ref{FC2}) formula (\ref%
{SumLamb}) taking $h(x)=xf(x,xr/a)/(a\omega )$. For this function the
residue term in (\ref{SumLamb}) vanishes. The part in the fermionic
condensate with the last integral on the right-hand side of (\ref{SumLamb})
vanishes in the limit $a\rightarrow \infty $, whereas the part with the
first integral does not depend on $a$. From here it follows that the latter
presents the fermionic condensate in the geometry of a cosmic string without
boundaries. This can also be seen by direct evaluation.

Indeed, when the cylindrical boundary is absent, the positive and negative
frequency eigenspinors are still given by formulae (\ref{psi+}) and (\ref%
{psi-}), where now the spectrum for $\lambda $ is continuous, $0\leqslant
\lambda <\infty $. The corresponding normalization coefficients are found
from the condition (\ref{normcond}) and have the form%
\begin{equation}
(C_{\sigma }^{(\pm )})^{-2}=2\pi \phi _{0}\frac{\kappa _{s}^{2}+1}{\lambda }%
(1+b_{s}^{(\pm )2}).  \label{CsigFree}
\end{equation}%
Substituting the eigenspinors into the mode-sum formula (\ref{FCmodesum}),
for the fermionic condensate in the geometry of a cosmic string without
boundaries we find%
\begin{eqnarray}
\langle 0|\bar{\psi}\psi |0\rangle _{\mathrm{s}} &=&-\frac{q}{8\pi ^{2}}%
\sum_{j=\pm 1/2,\pm 3/2,\cdots }\int_{-\infty }^{+\infty
}dk\,\int_{0}^{\infty }d\lambda \sum_{s=\pm 1}\frac{\lambda }{\omega }
\notag \\
&&\times \lbrack (m-s\sqrt{\lambda ^{2}+m^{2}})J_{\beta }^{2}(\lambda r)+(m+s%
\sqrt{\lambda ^{2}+m^{2}})J_{\beta +\epsilon _{j}}^{2}(\lambda r)].
\label{FC0}
\end{eqnarray}%
This coincides with the result obtained from the first term on the right of
formula (\ref{SumLamb}) applied to mode-sum (\ref{FC2}). Formula (\ref{FC0})
is further simplified by taking into account the expression for $\beta $ and
after the summation over $s$:%
\begin{equation}
\langle 0|\bar{\psi}\psi |0\rangle _{\mathrm{s}}=-\frac{qm}{\pi ^{2}}%
\sum_{j}\int_{0}^{\infty }dk\,\int_{0}^{\infty }d\lambda \frac{\lambda }{%
\omega }[J_{qj-1/2}^{2}(\lambda r)+J_{qj+1/2}^{2}(\lambda r)].  \label{FC0b}
\end{equation}%
Here and in what follows%
\begin{equation}
\sum_{j}=\sum_{j=1/2,3/2,\cdots }.  \label{SumjNot}
\end{equation}%
As it is seen from formula (\ref{FC0b}), for a massless field the fermionic
condensate vanishes in the boundary-free cosmic string spacetime. Since the
geometry outside of the string is flat, the renormalization of the fermionic
condensate given by (\ref{FC0b}) is done by subtracting the corresponding
quantity for the boundary-free Minkowski spacetime. The latter is obtained
from (\ref{FC0b}) taking in this formula $q=1$. Note that for the Minkowski
case the summation over $j$ is explicitly done by using the formula
\begin{equation}
\sum_{j}[J_{j-1/2}^{2}(\lambda r)+J_{j+1/2}^{2}(\lambda r)]=2%
\sideset{}{'}{\sum}_{n=0}^{\infty }J_{n}^{2}(\lambda r)=1,  \label{SumJ2}
\end{equation}%
where the prime on the sign of summation means that the $n=0$ term should be
halved. The renormalized fermionic condensate in the geometry of a cosmic
string without boundaries is further investigated in appendix \ref%
{sec:Appendix}.

From the discussion above it follows that, after the application of the
summation formula (\ref{SumLamb}), the part in the fermionic condensate with
the second integral on the right-hand side of this formula corresponds to
the VEV induced by the presence of the cylindrical shell. By using property (%
\ref{fprop}) and noting that under the change $s\rightarrow -s$ we have $%
F^{(+)}(x)\rightarrow F^{(-)}(x)$, $F^{(-)}(x)\rightarrow F^{(+)}(x)$, the
fermionic condensate in the geometry with the cylindrical shell is presented
in the decomposed form
\begin{equation}
\langle 0|\bar{\psi}\psi |0\rangle =\langle 0|\bar{\psi}\psi |0\rangle _{%
\mathrm{s}}+\langle \bar{\psi}\psi \rangle _{\mathrm{cyl}}.  \label{FC4}
\end{equation}%
Here the second term on the right-hand side,%
\begin{eqnarray}
\langle \bar{\psi}\psi \rangle _{\mathrm{cyl}} &=&\frac{q}{4\pi ^{3}a^{2}}%
\sum_{j=\pm 1/2,\pm 3/2,\cdots }\sum_{s}\int_{-\infty }^{+\infty }dk\int_{%
\sqrt{\mu ^{2}+a^{2}k^{2}}}^{\infty }\frac{dx\,x}{\sqrt{x^{2}-\mu
^{2}-a^{2}k^{2}}}  \notag \\
&&\frac{K_{\beta }^{(+)}(x)}{I_{\beta }^{(+)}(x)}[(\mu -is\sqrt{x^{2}-\mu
^{2}})I_{\beta }^{2}(xr/a)-(\mu +is\sqrt{x^{2}-\mu ^{2}})I_{\beta +\epsilon
_{j}}^{2}(xr/a)],  \label{FC3}
\end{eqnarray}%
is the part in the fermionic condensate induced by the cylindrical boundary.
The number of the integrations in this formula is reduced by using the
relation%
\begin{equation}
\int_{0}^{\infty }dk\int_{\sqrt{\mu ^{2}+a^{2}k^{2}}}^{\infty }\frac{%
dx\,xG(x)}{\sqrt{x^{2}-\mu ^{2}-a^{2}k^{2}}}=\frac{\pi }{2a}\int_{\mu
}^{\infty }duuG(u).  \label{IntFormula}
\end{equation}%
Further, redefining $s\rightarrow -s$ in the part of the summation over the
negative values $j$, it can be seen that the negative and positive values of
$j$ give the same contributions to the fermionic condensate. Finally, we
arrive to the following formula
\begin{equation}
\langle \bar{\psi}\psi \rangle _{\mathrm{cyl}}=\frac{q}{\pi ^{2}a^{3}}%
\sum_{j}\int_{\mu }^{\infty }dx\,x{\mathrm{Re}}\bigg[\frac{\bar{K}_{\beta
_{j}}(x)}{\bar{I}_{\beta _{j}}(x)}F_{\beta _{j}}(x,xr/a)\bigg],
\label{FCcyl}
\end{equation}%
where
\begin{equation}
\beta _{j}=qj-1/2,  \label{betaj}
\end{equation}%
and the function $F_{\beta _{j}}(x,y)$ is given by the expression
\begin{equation}
F_{\beta _{j}}(x,y)=(\mu -i\sqrt{x^{2}-\mu ^{2}})I_{\beta _{j}}^{2}(y)-(\mu
+i\sqrt{x^{2}-\mu ^{2}})I_{\beta _{j}}^{2}(y).  \label{Fbeta}
\end{equation}%
In (\ref{FCcyl}), for the modified Bessel functions we use the barred
notations%
\begin{eqnarray}
\bar{F}_{\beta }(x) &=&xF_{\beta }^{\prime }(x)+(\mu -i\sqrt{x^{2}-\mu ^{2}}%
-\beta )F_{\beta }(x)  \notag \\
&=&\eta _{F}xF_{\beta +1}(x)+(\mu -i\sqrt{x^{2}-\mu ^{2}})F_{\beta }(x),
\label{Fbarnot}
\end{eqnarray}%
with $F=I,K$, and $\eta _{I}=1$, $\eta _{K}=-1$.

The boundary induced part (\ref{FCcyl}) is finite for points away from the
cylindrical shell and the renormalization is necessary for the boundary-free
part, $\langle 0|\bar{\psi}\psi |0\rangle _{\mathrm{s}}$, only. Note that
the ratio in the integrand of Eq. (\ref{FCcyl}) may also be presented in the
form%
\begin{equation}
\frac{\bar{K}_{\beta }(x)}{\bar{I}_{\beta }(x)}=-\frac{W_{\beta }(x)-\mu +i%
\sqrt{x^{2}-\mu ^{2}}}{x^{2}[I_{\beta +1}^{2}(x)+I_{\beta }^{2}(x)]+2\mu
xI_{\beta }(x)I_{\beta +1}(x)},  \label{KbIb}
\end{equation}%
where%
\begin{equation}
W_{\beta }(x)=x^{2}\left[ I_{\beta +1}(x)K_{\beta +1}(x)-K_{\beta
}(x)I_{\beta }(x)\right] +2\mu xI_{\beta }(x)K_{\beta +1}(x).  \label{Wbeta}
\end{equation}%
For a massless fermionic field, from (\ref{FCcyl}) and (\ref{KbIb}) we find%
\begin{equation}
\langle \bar{\psi}\psi \rangle _{\mathrm{cyl}}=-\frac{q}{\pi ^{2}a^{3}}%
\sum_{j}\int_{0}^{\infty }dx\,x\frac{I_{\beta _{j}}^{2}(xr/a)+I_{\beta
_{j}+1}^{2}(xr/a)}{I_{\beta _{j}}^{2}(x)+I_{\beta _{j}+1}^{2}(x)}\mathrm{.}
\label{FCm0}
\end{equation}%
In this case the boundary-free part vanishes and the fermionic condensate is
always negative.

Now we turn to the investigation of the fermionic condensate given by Eq. (%
\ref{FCcyl}) in the asymptotic regions of the parameters. For large values
of the cylinder radius we use the asymptotic formulae for the modified
Bessel functions when the argument is large. By taking into account that the
main contribution into the integral in (\ref{FCcyl}) comes from the lower
limit of the integral, to the leading order we find%
\begin{equation}
\langle \bar{\psi}\psi \rangle _{\mathrm{cyl}}\approx -\frac{qm}{4\pi a^{2}}%
e^{-2am}\sum_{j}\sum_{\delta =\pm 1}(1+\delta qj)I_{qj-\delta /2}^{2}(mr),
\label{FClargea}
\end{equation}%
for $am\gg 1$. For a massless field, expanding the integrand in (\ref{FCm0})
we see that the main contribution is due to the term with $j=1/2$ and one has%
\begin{equation}
\langle \bar{\psi}\psi \rangle _{\mathrm{cyl}}\approx -\frac{2^{1-q}q}{\pi
^{2}a^{3}}\frac{(r/a)^{q-1}}{\Gamma ^{2}((q+1)/2)}\int_{0}^{\infty }dx\,%
\frac{x^{q}}{I_{(q-1)/2}^{2}(x)+I_{(q+1)/2}^{2}(x)},  \label{FClargeam0}
\end{equation}%
for $r\ll a$.

For points near the string, $r/a\ll 1$, the main contribution into the
boundary-induced VEV (\ref{FCcyl}) comes from the lowest mode $j=1/2$ and to
the leading order we find%
\begin{equation}
\langle \bar{\psi}\psi \rangle _{\mathrm{cyl}}\approx \frac{%
2^{1-q}q(r/a)^{q-1}}{\pi ^{2}a^{3}\Gamma ^{2}((q+1)/2)}\int_{\mu }^{\infty
}dx\,x^{q}{\mathrm{Re}}\bigg[\frac{\bar{K}_{(q-1)/2}(x)}{\bar{I}_{(q-1)/2}(x)%
}(\mu -i\sqrt{x^{2}-\mu ^{2}})\bigg].  \label{FConAxis}
\end{equation}%
This quantity is non-zero in the case of the cylindrical boundary in the
Minkowski bulk and vanishes for the cosmic string geometry with $q>1$. For a
massless field formula (\ref{FConAxis}) is reduced to Eq. (\ref{FClargeam0}%
). For points near the boundary the main contribution comes from large
values of $j$ and in this case we can use the uniform asymptotic expansions
for the modified Bessel functions for large values of the order (see, for
instance, \cite{hand}). In this way, to the leading order we have%
\begin{equation}
\langle \bar{\psi}\psi \rangle _{\mathrm{cyl}}\approx -\frac{1}{4\pi
^{2}(a-r)^{3}},  \label{FCnear}
\end{equation}%
for $(1-r/a)\ll 1$. As we see, the leading term does not depend on the
planar angle deficit and corresponds to the same one obtained for a
cylindrical boundary in the Minkowski bulk.

Now we consider the limit when the parameter $q$ is large which corresponds
to small values of $\phi _{0}$ and, hence, to a large planar angle deficit.
In this limit the order of the modified Bessel functions in (\ref{FCcyl}) is
large and we replace these functions by their uniform asymptotic expansions.
Assuming fixed value of the ratio $r/a$, the integral is estimated by the
Laplace method and to the leading order we have%
\begin{equation}
\langle \bar{\psi}\psi \rangle _{\mathrm{cyl}}\approx -\frac{q^{2}\exp
[-(1-(r/a)^{2})\mu /2]}{2\pi ^{2}r^{3}[1-(r/a)^{2}]}\left( \frac{r}{a}%
\right) ^{q},\;q\gg 1.  \label{FClargeq}
\end{equation}%
Hence, for large values of the angle deficit the fermionic condensate is
exponentially suppressed.

\section{Energy-momentum tensor}

\label{sec:EMT}

In this section we consider the VEV of the energy-momentum tensor of the
fermionic field inside a cylindrical shell on which the field satisfies the
boundary condition (\ref{BagCond}). This VEV can be evaluated by making use
of the mode-sum formula
\begin{equation}
\left\langle 0\left\vert T_{\mu \nu }\right\vert 0\right\rangle =\frac{i}{2}%
\sum_{\sigma }[\bar{\psi}_{\sigma }^{(-)}(x)\gamma _{(\mu }\nabla _{\nu
)}\psi _{\sigma }^{(-)}(x)-(\nabla _{(\mu }\bar{\psi}_{\sigma
}^{(-)}(x))\gamma _{\nu )}\psi _{\sigma }^{(-)}(x)]\ ,  \label{modesum}
\end{equation}%
with the negative frequency eigenspinors given by (\ref{psi-}). We can see
that the vacuum energy-momentum tensor is diagonal and the separate
components are given by the formulae (no summation over $\nu $)
\begin{equation}
\left\langle 0\left\vert T_{\nu }^{\nu }\right\vert 0\right\rangle =\frac{q}{%
8\pi ^{2}a^{3}}\sum_{\sigma }\frac{x^{3}T_{\beta }(x)}{a\omega }f_{\beta
}^{(\nu )}(x,xr/a)|_{x=\lambda _{\beta ,l}},  \label{EMT}
\end{equation}%
where the following notations were introduced%
\begin{eqnarray}
f_{\beta }^{(0)}(x,y) &=&-\frac{a^{2}\omega ^{2}}{x^{2}}\bigg[\bigg(1-\frac{%
s\mu }{\sqrt{x^{2}+\mu ^{2}}}\bigg)J_{\beta }^{2}(y)+\bigg(1+\frac{s\mu }{%
\sqrt{x^{2}+\mu ^{2}}}\bigg)J_{\beta +\epsilon _{j}}^{2}(y)\bigg],  \notag \\
f_{\beta }^{(1)}(x,y) &=&J_{\beta }^{2}(y)+J_{\beta +\epsilon _{j}}^{2}(y)-%
\frac{2\beta +\epsilon _{j}}{y}J_{\beta }(y)J_{\beta +\epsilon _{j}}(y),
\label{fbetamu} \\
f_{\beta }^{(2)}(x,y) &=&\frac{2\beta +\epsilon _{j}}{y}J_{\beta
}(y)J_{\beta +\epsilon _{j}}(y),\;\,f_{\beta }^{(3)}(x,y)=(k^{2}/\omega
^{2})f_{\beta }^{(0)}(x,y).  \notag
\end{eqnarray}%
The other notations in (\ref{EMT}) are the same as in (\ref{FC2}). The VEV
given by (\ref{EMT}) is divergent and, as in the case of the fermionic
condensate, it is assumed that a cutoff function is present. Now, we can
explicitly verify that the VEVs (\ref{EMT}) satisfy the trace relation
\begin{equation*}
\left\langle 0\left\vert T_{\nu }^{\nu }\right\vert 0\right\rangle =m\langle
0|\bar{\psi}\psi |0\rangle .
\end{equation*}

In order to extract from the VEV of the energy-momentum tensor the part
corresponding to the geometry of a string without boundaries, we apply to
the series over $l$ in (\ref{EMT}) the summation formula (\ref{SumLamb}) with%
\begin{equation}
h(x)=\frac{x^{3}}{a\omega }f_{\beta }^{(\nu )}(x,xr/a).  \label{hxEMT}
\end{equation}%
In a way similar to that used in the case of the fermionic condensate, the
VEV can be written in the decomposed form:%
\begin{equation}
\left\langle 0\left\vert T_{\mu }^{\nu }\right\vert 0\right\rangle
=\left\langle 0\left\vert T_{\mu }^{\nu }\right\vert 0\right\rangle _{%
\mathrm{s}}+\left\langle T_{\mu }^{\nu }\right\rangle _{\mathrm{cyl}},
\label{EMT1}
\end{equation}%
where $\left\langle 0\left\vert T_{\mu }^{\nu }\right\vert 0\right\rangle _{%
\mathrm{s}}$ is the fermionic energy-momentum tensor in the geometry of a
cosmic string when the cylindrical shell is absent. The second term on the
right-hand side of formula (\ref{EMT1}) is induced by the cylindrical shell
and is given by the formula (no summation over $\nu $)%
\begin{equation}
\left\langle T_{\nu }^{\nu }\right\rangle _{\mathrm{cyl}}=\frac{q}{\pi
^{2}a^{4}}\sum_{j}\int_{\mu }^{\infty }dx\,x^{3}{\mathrm{Re}}\bigg[\frac{%
\bar{K}_{\beta _{j}}(x)}{\bar{I}_{\beta _{j}}(x)}F_{\beta _{j}}^{(\nu
)}(x,xr/a)\bigg],  \label{EMTcyl}
\end{equation}%
where $\beta _{j}$ is defined by Eq. (\ref{betaj}). In this formula we have
introduced the notations%
\begin{eqnarray}
F_{\beta _{j}}^{(0)}(x,y) &=&\frac{\mu ^{2}/x^{2}-1}{2}\sum_{\delta =\pm
1}\delta \bigg(1+\frac{\delta i\mu }{\sqrt{x^{2}-\mu ^{2}}}\bigg)%
I_{qj-\delta /2}^{2}(y),  \notag \\
F_{\beta _{j}}^{(1)}(x,y) &=&I_{\beta _{j}}^{2}(y)-I_{\beta
_{j}+1}^{2}(y)-(2qj/y)I_{\beta _{j}}(y)I_{\beta _{j}+1}(y),  \label{Fbetamu}
\\
F_{\beta _{j}}^{(2)}(x,y) &=&(2qj/y)I_{\beta _{j}}(y)I_{\beta _{j}+1}(y),
\notag
\end{eqnarray}%
and $F_{\beta }^{(3)}(x,y)=F_{\beta }^{(0)}(x,y)$. Note that the function
for the radial component is also presented in the form%
\begin{equation}
F_{\beta _{j}}^{(1)}(x,y)=I_{\beta _{j}}(y)I_{\beta _{j}+1}^{\prime
}(y)-I_{\beta _{j}+1}(y)I_{\beta _{j}}^{\prime }(y).  \label{Fbet1}
\end{equation}%
As we see, the vacuum stress along the axis of the string is equal to the
energy density. Of course, this property is a direct consequence of the bust
invariance along this axis.

By using Eq. (\ref{KbIb}), we may write the expressions for the components
of the energy-momentum tensor in more explicit form given by
\begin{eqnarray}
\left\langle T_{0}^{0}\right\rangle _{\mathrm{cyl}} &=&\frac{q}{2\pi
^{2}a^{4}}\sum_{j}\int_{\mu }^{\infty }dx\,x(1-\mu ^{2}/x^{2})  \notag \\
&&\times \frac{W_{\beta _{j}}(x)[I_{\beta _{j}}^{2}(xr/a)-I_{\beta
_{j}+1}^{2}(xr/a)]-2\mu I_{\beta _{j}}^{2}(xr/a)}{I_{\beta
_{j}+1}^{2}(x)+I_{\beta _{j}}^{2}(x)+2(\mu /x)I_{\beta _{j}}(x)I_{\beta
_{j}+1}(x)},  \label{T00int}
\end{eqnarray}%
for the energy density and by (no summation over $\nu $)%
\begin{equation}
\left\langle T_{\nu }^{\nu }\right\rangle _{\mathrm{cyl}}=-\frac{q}{\pi
^{2}a^{4}}\sum_{j}\int_{\mu }^{\infty }dx\,\frac{x[W_{\beta _{j}}(x)-\mu
]F_{\beta _{j}}^{(\nu )}(x,xr/a)}{I_{\beta _{j}+1}^{2}(x)+I_{\beta
_{j}}^{2}(x)+2(\mu /x)I_{\beta _{j}}(x)I_{\beta _{j}+1}(x)},  \label{Tnuint}
\end{equation}%
for the radial and azimuthal stresses, $\nu =1,2$.

The VEV of the fermionic energy-momentum tensor in the geometry of a cosmic
string when the cylindrical shell is absent, corresponds to the first term
on the right-hand side of summation formula (\ref{SumLamb}). For this VEV we
have the mode-sum (no summation over $\nu $)%
\begin{equation}
\left\langle 0\left\vert T_{\nu }^{\nu }\right\vert 0\right\rangle _{\mathrm{%
s}}=\frac{q}{8\pi ^{2}}\sum_{j=\pm 1/2,\pm 3/2,\cdots }\int_{-\infty
}^{+\infty }dk\,\int_{0}^{\infty }d\lambda \sum_{s=\pm 1}\frac{\lambda ^{3}}{%
\omega }f_{\beta }^{(\nu )}(\lambda a,\lambda r).  \label{EMTs}
\end{equation}%
The summation over $s$ in this formula is done explicitly and noting that
the negative and positive values of $j$ give the same contributions, we find%
\begin{equation}
\left\langle 0\left\vert T_{\nu }^{\nu }\right\vert 0\right\rangle _{\mathrm{%
s}}=\frac{q}{\pi ^{2}}\sum_{j}\int_{0}^{\infty }dk\,\int_{0}^{\infty
}d\lambda \frac{\lambda ^{3}}{\omega }g_{\beta _{j}}^{(\nu )}(\lambda
,\lambda r),  \label{EMTs1}
\end{equation}%
where now%
\begin{eqnarray}
g_{\beta _{j}}^{(0)}(\lambda ,y) &=&-\frac{\omega ^{2}}{\lambda ^{2}}%
[J_{\beta _{j}}^{2}(y)+J_{\beta _{j}+1}^{2}(y)],  \notag \\
g_{\beta _{j}}^{(1)}(\lambda ,y) &=&J_{\beta _{j}}^{2}(y)+J_{\beta
_{j}+1}^{2}(y)-\frac{2qj}{y}J_{\beta _{j}}(y)J_{\beta _{j}+1}(y),
\label{gmu} \\
g_{\beta _{j}}^{(2)}(\lambda ,y) &=&\frac{2qj}{y}J_{\beta _{j}}(y)J_{\beta
_{j}+1}(y),\;g_{\beta _{j}}^{(3)}(\lambda ,y)=(k^{2}/\omega ^{2})g_{\beta
_{j}}^{(0)}(\lambda ,y).  \notag
\end{eqnarray}%
As in the case of the fermionic condensate, the VEV (\ref{EMTs1}) is
renormalized by subtracting the corresponding VEV in the Minkowski spacetime
without boundaries. The boundary-free renormalized VEV\ of the
energy-momentum tensor for a massive fermionic field is further investigated
in appendix \ref{sec:Appendix}.

\begin{figure}[tbph]
\begin{center}
\begin{tabular}{cc}
\epsfig{figure=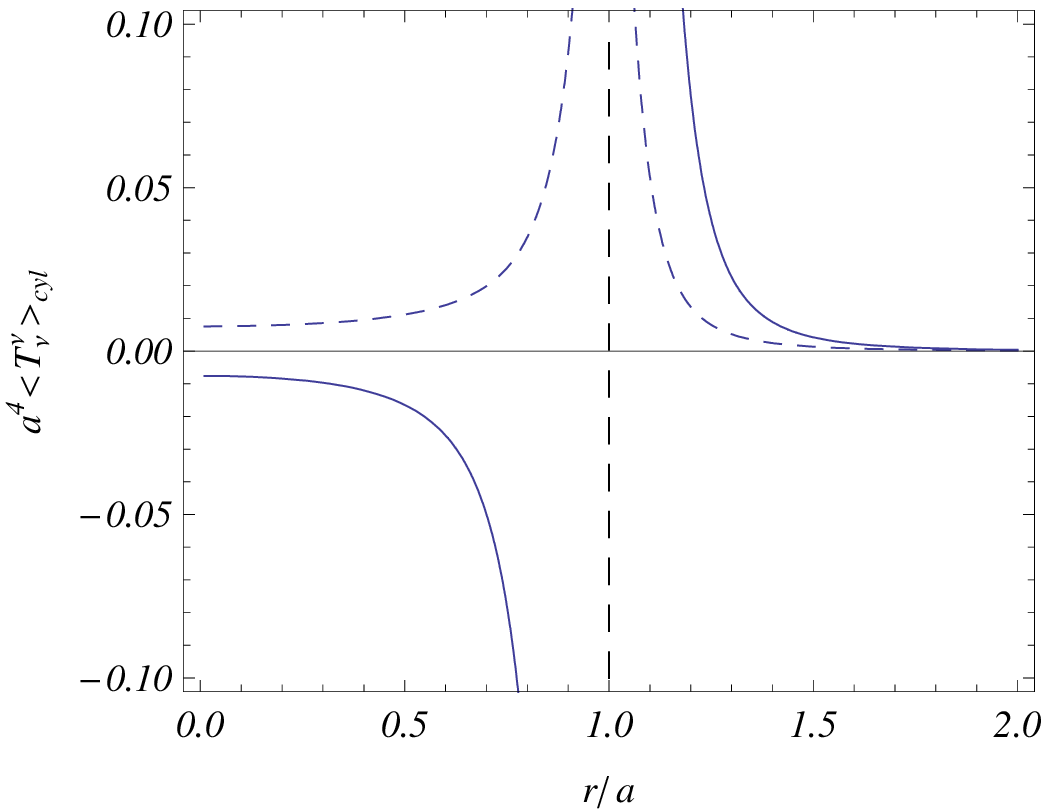,width=7.cm,height=6.cm} & \quad %
\epsfig{figure=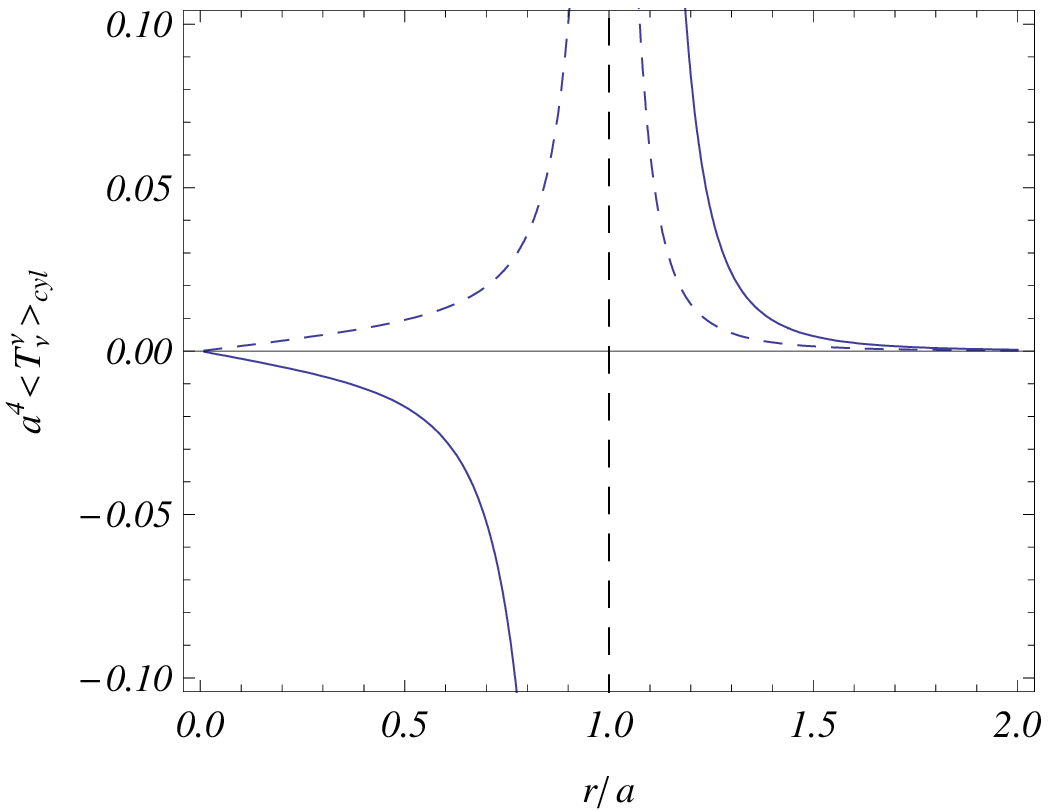,width=7.cm,height=6.cm}%
\end{tabular}%
\end{center}
\caption{Boundary induced parts in the VEVs of the energy density, $%
a^{4}\langle T_{0}^{0}\rangle _{\mathrm{cyl}}$ (full curves), and radial
stress, $a^{4}\langle T_{1}^{1}\rangle _{\mathrm{cyl}}$ (dashed curves), for
a massless fermionic field as functions of the radial coordinate. The left
panel is plotted for $q=1$ (Minkowski spacetime) and for the right panel $%
q=2 $.}
\label{fig1}
\end{figure}

It can be explicitly verified that the boundary-induced parts satisfy the
trace relation $\langle T_{\nu }^{\nu }\rangle _{\mathrm{cyl}}=m\langle \bar{%
\psi}\psi \rangle _{\mathrm{cyl}}$ and the covariant conservation equation $%
\nabla _{\nu }\langle T_{\mu }^{\nu }\rangle _{\mathrm{cyl}}=0$. For the
geometry under consideration the latter is reduced to the single equation%
\begin{equation}
\partial _{r}[r\left\langle T_{1}^{1}\right\rangle _{\mathrm{cyl}%
}]=\left\langle T_{2}^{2}\right\rangle _{\mathrm{cyl}}.  \label{conteq}
\end{equation}%
This equation is easily checked by using the relation $\partial
_{y}(yF_{\beta }^{(1)}(x,y))=F_{\beta }^{(2)}(x,y)$ between the radial and
azimuthal functions in (\ref{Fbetamu}). In the case of a massless fermionic
field, for the VEV induced by the cylindrical boundary from (\ref{T00int})
and (\ref{Tnuint}) we have (no summation over $\nu $)
\begin{equation}
\left\langle T_{\nu }^{\nu }\right\rangle _{\mathrm{cyl}}=\frac{q}{\pi
^{2}a^{4}}\sum_{j}\int_{0}^{\infty }dx\,x^{3}\frac{I_{\beta _{j}}(x)K_{\beta
_{j}}(x)-I_{\beta _{j}+1}(x)K_{\beta _{j}+1}(x)}{I_{\beta
_{j}}^{2}(x)+I_{\beta _{j}+1}^{2}(x)}F_{\beta _{j}}^{(0,\nu )}(xr/a),
\label{EMTm0}
\end{equation}%
where $F_{\beta }^{(0,\nu )}(y)=F_{\beta }^{(\nu )}(x,y)$\ for $\nu =1,2$,
and the corresponding function for the energy density is defined as%
\begin{equation}
F_{\beta _{j}}^{(0,0)}(y)=-\frac{1}{2}[I_{\beta _{j}}^{2}(y)-I_{\beta
_{j}+1}^{2}(y)].  \label{Fbeta0m0}
\end{equation}%
In the special case with $q=1$, from the formulae given above we obtain the
fermionic Casimir densities for a cylindrical boundary in the Minkowski
spacetime. \ On the left panel of figutre \ref{fig1} we have plotted the
corresponding VEVs for the energy density and radial stress as functions of
the radial coordinate for a massless fermionic field. On the right panel the
boundary induced parts are presented in the geometry of a cosmic string with
the parameter $q=2$.

For large values of the cylinder radius the main contribution into the
integral in (\ref{EMTcyl}) comes from the lower limit and in the leading
order we have (no summation over $\nu $)%
\begin{eqnarray}
\left\langle T_{0}^{0}\right\rangle _{\mathrm{cyl}} &\approx &-\frac{qm^{2}}{%
8\pi a^{2}}e^{-2am}\sum_{j}\sum_{\delta =\pm 1}I_{qj-\delta /2}^{2}(mr),
\notag \\
\left\langle T_{\nu }^{\nu }\right\rangle _{\mathrm{cyl}} &\approx &-\frac{%
q^{2}m^{2}}{4\pi a^{2}}e^{-2am}\sum_{j}qjF_{\beta _{j}}^{(\nu )}(ma,mr),
\label{EMTlargea}
\end{eqnarray}%
where $\nu =1,2$ and $am\gg 1$. For a massless fermionic field the main
contribution comes from $j=1/2$ term and from (\ref{EMTm0}) one finds%
\begin{equation}
\left\langle T_{0}^{0}\right\rangle _{\mathrm{cyl}}\approx -\frac{%
2^{-q}q(r/a)^{q-1}}{\pi ^{2}a^{4}\Gamma ^{2}((q+1)/2)}\int_{0}^{\infty
}dx\,x^{q+2}\frac{I_{(q-1)/2}(x)K_{(q-1)/2}(x)-I_{(q+1)/2}(x)K_{(q+1)/2}(x)}{%
I_{(q-1)/2}^{2}(x)+I_{(q+1)/2}^{2}(x)},  \label{T00largeam0}
\end{equation}%
for the energy density and
\begin{equation}
\left\langle T_{1}^{1}\right\rangle _{\mathrm{cyl}}\approx \frac{-2}{q+1}%
\left\langle T_{0}^{0}\right\rangle _{\mathrm{cyl}},\;\left\langle
T_{2}^{2}\right\rangle _{\mathrm{cyl}}\approx \frac{-2q}{q+1}\left\langle
T_{0}^{0}\right\rangle _{\mathrm{cyl}},  \label{T11largeam0}
\end{equation}%
for the radial and azimuthal stresses.

Now we consider the behavior of the boundary induced VEV of the
energy-momentum tensor near the string and for points near the boundary. In
the first case one has $r/a\ll 1$ and the main contribution in (\ref{EMTcyl}%
) comes from the lowest mode $j=1/2$. By using the formulae for the modified
Bessel functions for small values of the argument, to the leading order one
finds (no summation over $\nu $)%
\begin{equation}
\langle T_{\nu }^{\nu }\rangle _{\mathrm{cyl}}\approx \frac{q(r/2a)^{q-1}}{%
\pi ^{2}a^{4}\Gamma ^{2}((q+1)/2)}\int_{\mu }^{\infty }dx\,x^{q+2}{\mathrm{Re%
}}\bigg[\frac{\bar{K}_{(q-1)/2}(x)}{\bar{I}_{(q-1)/2}(x)}F^{(\nu )}(x)\bigg],
\label{EMTcylnearStr}
\end{equation}%
where%
\begin{eqnarray}
F^{(0)}(x) &=&\frac{\mu ^{2}/x^{2}-1}{2}(1+i\mu /\sqrt{x^{2}-\mu ^{2}}),
\notag \\
F^{(2)}(x) &=&qF^{(1)}(x)=q/(q+1).  \label{Fmu0}
\end{eqnarray}%
For a massless fermionic field this formula is reduced to the results given
by Eqs. (\ref{T00largeam0}) and (\ref{T11largeam0}).

For points near the cylindrical boundary, we replace the modified Bessel
functions by the corresponding uniform asymptotic expansions for large
values of the order. Unlike to the case of the fermionic condensate here the
leading terms in the VEVs of the energy-momentum tensor vanish and it is
necessary to take the next terms in the asymptotic expansions. In
particular, for the function appearing in the integrands of the vacuum
stresses we have%
\begin{equation}
{\mathrm{Re}}\bigg[\frac{\bar{K}_{\beta }(\beta x)}{\bar{I}_{\beta }(\beta x)%
}\bigg]\approx \frac{1-t^{2}-2\beta }{1-t}\frac{\pi t^{2}}{2\beta }%
e^{-2\beta \eta (x)},  \label{ReKAs}
\end{equation}%
where%
\begin{equation}
t=\frac{1}{\sqrt{1+x^{2}}},\;\eta (x)=\sqrt{1+x^{2}}+\ln \frac{x}{1+\sqrt{%
1+x^{2}}}.  \label{teta}
\end{equation}%
Substituting this into the expression for the azimuthal stress, to the
leading order we find%
\begin{equation}
\left\langle T_{2}^{2}\right\rangle _{\mathrm{cyl}}\approx \frac{1-5\mu }{%
60\pi ^{2}a(a-r)^{3}},\;(1-r/a)\ll 1.  \label{T22NearCyl}
\end{equation}%
The corresponding expressions for the radial stress and the energy density
are found from the conservation equation and the trace relation by using the
result (\ref{FCnear}) for the fermionic condensate. In this way we obtain
the following formulae:%
\begin{equation}
\langle T_{0}^{0}\rangle _{\mathrm{cyl}}\approx -\frac{1+10\mu }{120\pi
^{2}a(a-r)^{3}},\;\left\langle T_{1}^{1}\right\rangle _{\mathrm{cyl}}\approx
\frac{1-5\mu }{120\pi ^{2}a^{2}(a-r)^{2}}.  \label{T00nearCyl}
\end{equation}%
As in the case of the fermionic condensate, the leading terms in the VEVs of
the energy-momentum tensor components do not depend on the planar angle
deficit and are the same as for a cylindrical boundary in the Minkowski
bulk. Note that, in dependence of the parameter $\mu $, the vacuum stresses
near the boundary can be either positive or negative, whereas the energy
density remains always negative. As the boundary-free part is finite
everywhere outside the string axis, for points near the boundary the total
VEV is dominated by the boundary induced part.

\begin{figure}[tbph]
\begin{center}
\begin{tabular}{cc}
\epsfig{figure=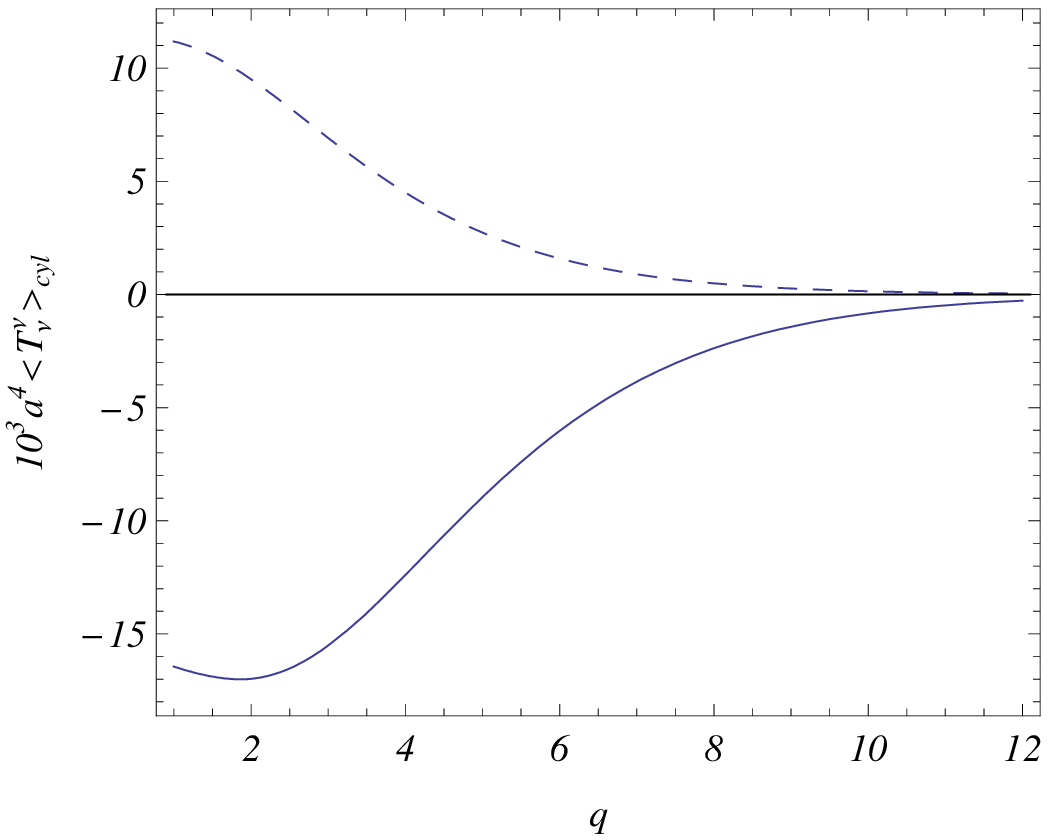,width=7.cm,height=6.cm} & \quad %
\epsfig{figure=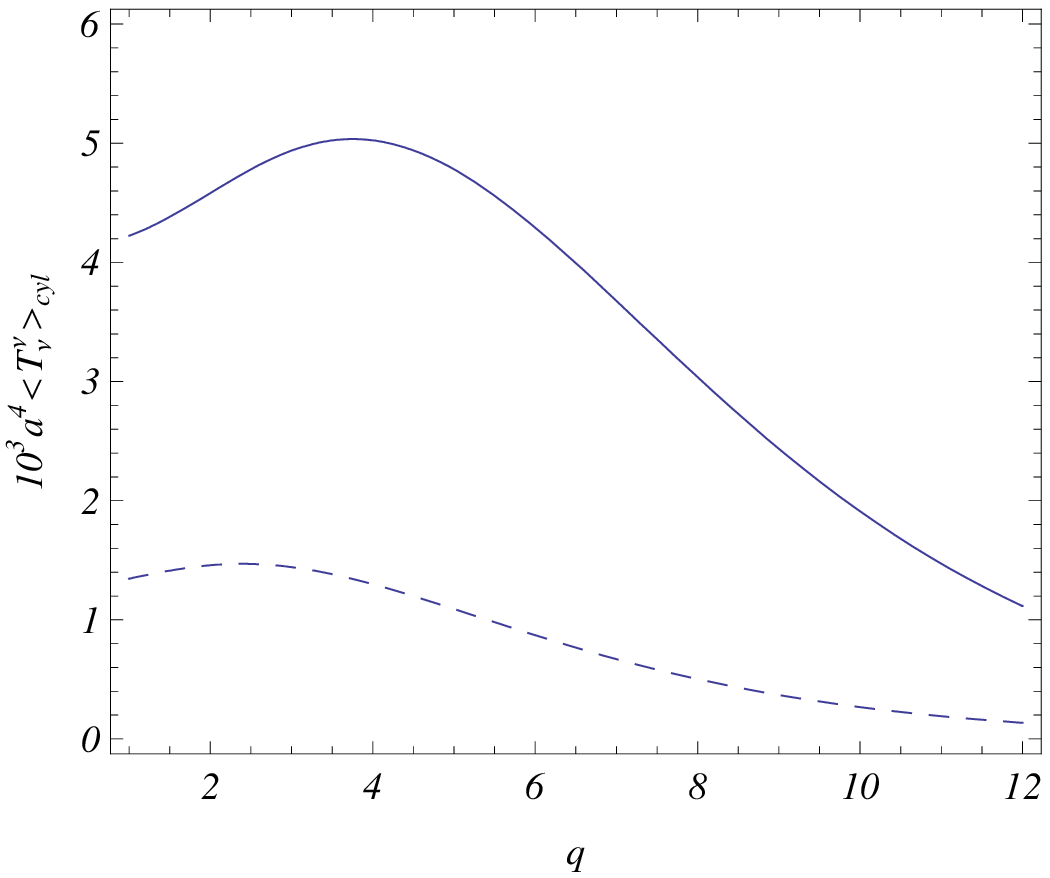,width=7.cm,height=6.cm}%
\end{tabular}%
\end{center}
\caption{Boundary induced parts in the VEVs of the energy density, $%
a^{4}\langle T_{0}^{0}\rangle _{\mathrm{cyl}}$ (full curves), and radial
stress, $a^{4}\langle T_{1}^{1}\rangle _{\mathrm{cyl}}$ (dashed curves), for
a massless fermionic field as functions of the parameter $q$. The left panel
corresponds to the interior region with $r/a=0.5$ and the right panel
corresponds to the exterior region with $r/a=1.5$.}
\label{fig2}
\end{figure}

Similar to the case of the fermionic condensate, it can be seen that for
large values of the parameter $q$ the boundary induced part in the VEV of
the energy-momentum tensor is suppressed by the factor $(r/a)^{q}$. The
dependence of the boundary induced VEVs on the parameter $q$ is presented in
figure \ref{fig2} for a massless fermionic field. On the left (right) panel
the energy density and radial stresses are plotted for the value of the
ratio $r/a=0.5$ ($r/a=1.5$).

\begin{figure}[tbph]
\begin{center}
\begin{tabular}{cc}
\epsfig{figure=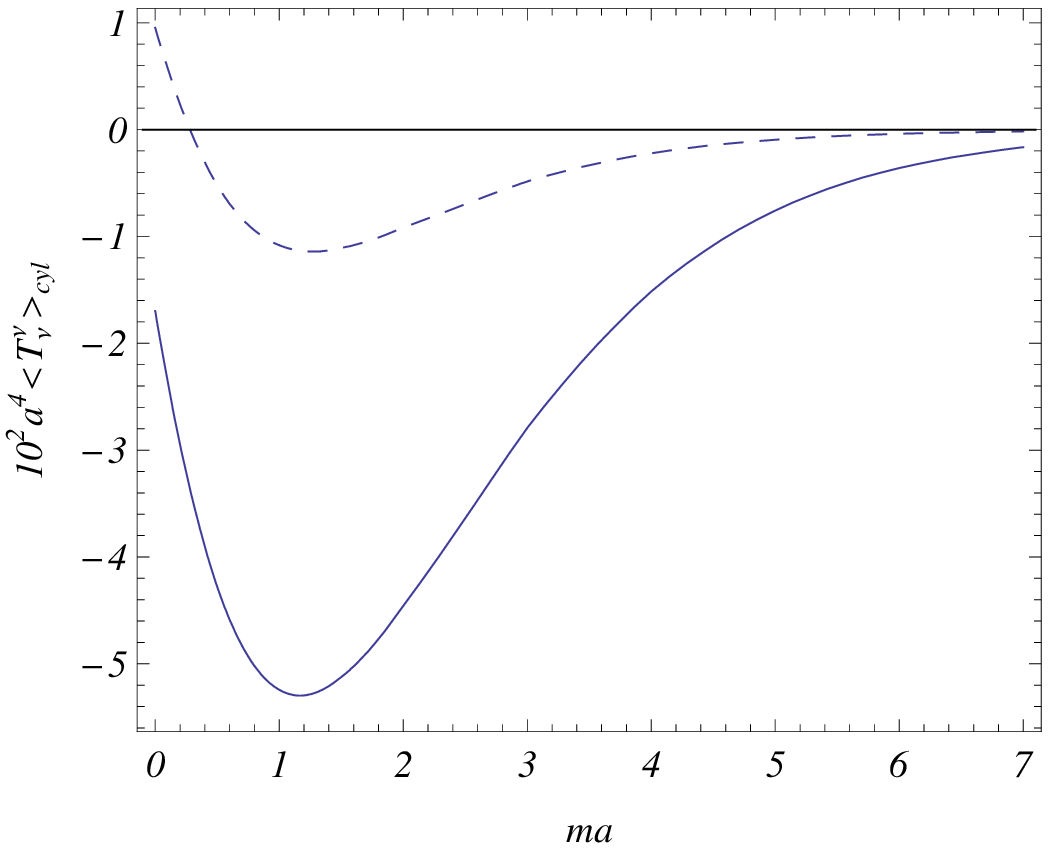,width=7.cm,height=6.cm} & \quad %
\epsfig{figure=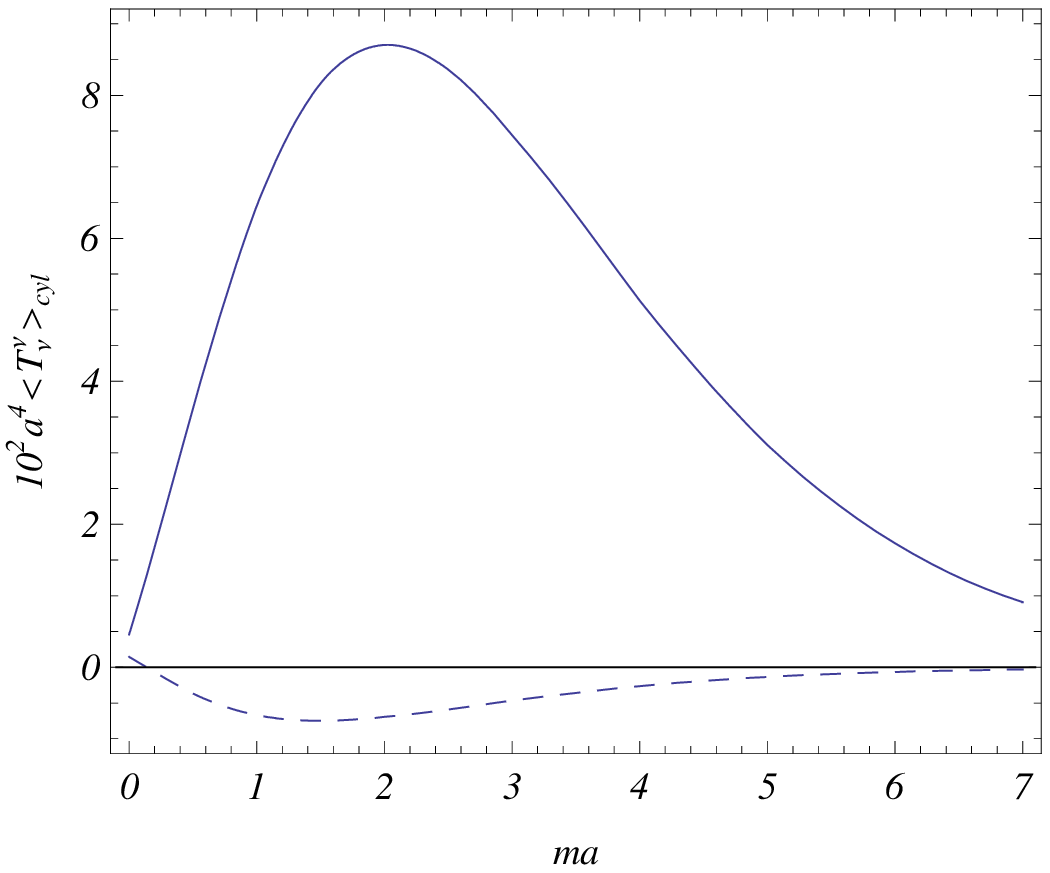,width=7.cm,height=6.cm}%
\end{tabular}%
\end{center}
\caption{Boundary induced parts in the VEVs of the energy density, $a^4
\langle T_{0}^{0}\rangle _{\mathrm{cyl}}$ (full curves), and radial stress, $%
a^4 \langle T_{1}^{1}\rangle _{\mathrm{cyl}}$ (dashed curves), as functions
of the parameter $ma$ for $q=2$. The left panel corresponds to the interior
region with $r/a=0.5$ and the right panel corresponds to the exterior region
with $r/a=1.5$.}
\label{fig3}
\end{figure}

In figure \ref{fig3} we give the boundary induced parts in the energy
density and the radial stress in the geometry of a cosmic string with $q=2$
as functions of the mass. The graphs on the left panel are for the interior
quantities at $r/a=0.5$ and the graphs on the right panel are for the
exterior ones evaluated at $r/a=1.5$. As it is seen, we have a non-trivial
dependence on the mass and in the case of a massive field the polarization
effects induced by the boundary can be stronger than for a massless field.

\section{Vacuum densities in the exterior region}

\label{sec:Exterior}

\subsection{Eigenspinors}

In this section we consider the fermionic condensate and the VEV of the
energy-momentum tensor in the region outside the cylindrical shell. As in
the interior case this can be done by the direct mode summation. The
corresponding eigenspinors have the form (\ref{psi+}) and (\ref{psi-}) with
the difference that now, instead of the Bessel functions $J_{\beta
_{1,2}}(\lambda r)$, the linear combinations of the functions $J_{\beta
_{1,2}}(\lambda r)$ and $Y_{\beta _{1,2}}(\lambda r)$ should be taken, with $%
Y_{\nu }(x)$ being the Neumann function. The ratio of the coefficients in
this combination is determined from the boundary condition (\ref{BagCond})
imposed on the cylindrical surface. In this way for the positive and
negative frequency eigenspinors we have the expressions
\begin{equation}
\psi _{\sigma }^{(\pm )}=C_{\sigma }^{(\pm )}\left(
\begin{array}{c}
Z_{\beta _{\pm }}(\lambda a,\lambda r) \\
i\kappa _{s}\epsilon _{n_{\pm }}b_{s}^{(\pm )}Z_{\beta _{\mp }}(\lambda
a,\lambda r)e^{iq\phi } \\
\kappa _{s}Z_{\beta _{\pm }}(\lambda a,\lambda r) \\
-i\epsilon _{n_{\pm }}b_{s}^{(\pm )}Z_{\beta _{\mp }}(\lambda a,\lambda
r)e^{iq\phi }%
\end{array}%
\right) \exp \left[ \pm i\left( q(j\mp 1/2)\phi +kz-\omega t\right) \right] ,
\label{psiext}
\end{equation}%
where the function $Z_{\nu }(x,y)$ is defined by the formula%
\begin{equation}
Z_{\beta _{\pm }}(x,y)=\tilde{Y}_{\beta }(x)J_{\beta _{\pm }}(y)-\tilde{J}%
_{\beta }(x)Y_{\beta _{\pm }}(y),  \label{Zbet}
\end{equation}%
and%
\begin{equation}
\beta _{+}=\beta _{1}=\beta ,\;\beta _{-}=\beta _{2},  \label{betplmin}
\end{equation}%
with $\beta _{1,2}$ given by relations (\ref{bet12a}). Here the notation $%
\tilde{Y}_{\beta }(x)$ is defined by (\ref{ftilde}) with the replacement $%
J\rightarrow Y$.

The eigenspinors are orthonormalized by condition (\ref{normcond}), where
now the radial integration goes over the exterior region. The eigenvalues
for $\lambda $ are continuous and on the right-hand side of the
normalization condition we have $\delta (\lambda -\lambda ^{\prime })$.
Since the radial integral diverges for $\lambda ^{\prime }=\lambda $, the
main contribution to this integral comes from large values $r$ and we cane
replace the Bessel and Neumann functions with the arguments $\lambda r$, by
the corresponding asymptotic expressions. In this way, for the normalization
coefficients in (\ref{psiext}) we find:%
\begin{equation}
(C_{\sigma }^{(\pm )})^{-2}=2\pi \phi _{0}(\kappa _{s}^{2}+1)(1+b_{s}^{(\pm
)2})\frac{1}{\lambda }\left[ \tilde{J}_{\beta }^{2}(\lambda a)+\tilde{Y}%
_{\beta }^{2}(\lambda a)\right] ,  \label{Csigext}
\end{equation}%
where $\kappa _{s}$ and $b_{s}^{(\pm )}$ are defined by Eqs. (\ref{kappas})
and (\ref{bsplmin}).

\subsection{Fermionic condensate}

Substituting the eigenspinors (\ref{psiext}) into the mode-sum formula (\ref%
{FCmodesum}), for the fermionic condensate in the region outside the
cylindrical shell we obtain the formula%
\begin{eqnarray}
\langle 0|\bar{\psi}\psi |0\rangle &=&\frac{q}{8\pi ^{2}}\sum_{\sigma }\frac{%
\lambda }{\omega \left[ \bar{J}_{\beta }^{2}(\lambda a)+\bar{Y}_{\beta
}^{2}(\lambda a)\right] }  \notag \\
&&\times \lbrack (-m+s\sqrt{\lambda ^{2}+m^{2}})Z_{\beta }^{2}(\lambda
a,\lambda r)-(m+s\sqrt{\lambda ^{2}+m^{2}})Z_{\beta +\sigma
_{j}}^{2}(\lambda a,\lambda r)],  \label{FCext}
\end{eqnarray}%
where%
\begin{equation}
\sum_{\sigma }=\sum_{j=\pm 1/2,\pm 3/2,\cdots }\int_{-\infty }^{+\infty
}dk\,\int_{0}^{\infty }d\lambda \sum_{s=\pm 1}\,.  \label{sigSum}
\end{equation}%
In order to extract from the VEV (\ref{FCext}) the part induced by the
cylindrical shell, we subtract the fermionic condensate for the geometry of
a string without the shell. As it has been shown before, the latter is given
by formula (\ref{FC0}). For further evaluation of the difference, we use the
identities%
\begin{equation}
\frac{Z_{\beta +\sigma _{j}}^{2}(\lambda a,\lambda r)}{\tilde{J}_{\beta
}^{2}(\lambda a)+\tilde{Y}_{\beta }^{2}(\lambda a)}-J_{\beta +\sigma
_{j}}^{2}(\lambda r)=-\frac{1}{2}\sum_{l=1,2}\frac{\tilde{J}_{\beta
}(\lambda a)}{\tilde{H}_{\beta }^{(l)}(\lambda a)}H_{\beta +\sigma
_{j}}^{(l)2}(\lambda r),  \label{Ident}
\end{equation}%
where $\sigma _{j}=0$ or $\epsilon _{j}$, and $H_{\nu }^{(1,2)}(x)$ are the
Hankel functions. As a result for the fermionic condensate we obtain
\begin{eqnarray}
\langle 0|\bar{\psi}\psi |0\rangle &=&\langle 0|\bar{\psi}\psi |0\rangle _{%
\mathrm{s}}+\frac{q}{16\pi ^{2}}\sum_{\sigma }\sum_{l=1,2}\frac{\lambda }{%
\omega }\frac{\tilde{J}_{\beta }(\lambda a)}{\tilde{H}_{\beta
}^{(l)}(\lambda a)}  \notag \\
&&\times \lbrack (m-s\sqrt{\lambda ^{2}+m^{2}})H_{\beta }^{(l)2}(\lambda
r)+(m+s\sqrt{\lambda ^{2}+m^{2}})H_{\beta +\sigma _{j}}^{(l)2}(\lambda r)].
\label{FCext1}
\end{eqnarray}

Now, in the complex plane $\lambda $ we rotate the integration contour in
the integral over $\lambda $ on the right-hand side of formula (\ref{FCext1}%
) by the angle $\pi /2$ for $l=1$ term and by the angle $-\pi /2$ for $l=2$
term. By using the symmetry properties of the integrands, it can be seen
that the parts of the integrals over $(0,i\sqrt{k^{2}+m^{2}})$ and $(0,-i%
\sqrt{k^{2}+m^{2}})$ are cancelled. The number of the remained integrations
is reduced by using formula (\ref{IntFormula}). In this way, introducing the
modified Bessel functions, we present the fermionic condensate in the
decomposed form (\ref{FC4}), where the part induced by the cylindrical
boundary in the region $r>a$ is given by the expression%
\begin{equation}
\langle \bar{\psi}\psi \rangle _{\mathrm{cyl}}=\frac{q}{\pi ^{2}a^{3}}%
\sum_{j}\int_{\mu }^{\infty }dx\,x{\mathrm{Re}}\bigg[\frac{\bar{I}_{\beta
_{j}}(x)}{\bar{K}_{\beta _{j}}(x)}F_{\beta _{j}}^{\mathrm{(ex)}}(x,xr/a)%
\bigg],  \label{FCcylext}
\end{equation}%
where $\beta _{j}$ is defined by Eq. (\ref{betaj}) and we are using notation%
\begin{equation}
F_{\beta _{j}}^{\mathrm{(ex)}}(x,y)=(\mu -i\sqrt{x^{2}-\mu ^{2}})K_{\beta
_{j}}^{2}(y)-(\mu +i\sqrt{x^{2}-\mu ^{2}})K_{\beta _{j}}^{2}(y).
\label{Fext}
\end{equation}%
Note that the ratio of the combinations of the modified Bessel functions in (%
\ref{FCcylext}) can be written in the form%
\begin{equation}
\frac{\bar{I}_{\beta }(x)}{\bar{K}_{\beta }(x)}=-\frac{W_{\beta }(x)-\mu -i%
\sqrt{x^{2}-\mu ^{2}}}{x^{2}[K_{\beta +1}^{2}(x)+K_{\beta }^{2}(x)]-2\mu
xK_{\beta }(x)K_{\beta +1}(x)},  \label{IbKb}
\end{equation}%
where the function $W_{\beta }(x)$ is defined by Eq. (\ref{Wbeta}). For a
massless fermionic field one has the formula%
\begin{equation}
\langle \bar{\psi}\psi \rangle _{\mathrm{cyl}}=\frac{q}{\pi ^{2}a^{3}}%
\sum_{j}\int_{0}^{\infty }dx\,x\frac{K_{\beta _{j}}^{2}(xr/a)+K_{\beta
_{j}+1}^{2}(xr/a)}{K_{\beta _{j}}^{2}(x)+K_{\beta _{j}+1}^{2}(x)},
\label{FCcylextm0}
\end{equation}%
and the fermionic condensate is positive.

Let us consider the behavior of the fermionic condensate in asymptotic
regions of the parameter. In the limit $a\rightarrow 0$ with fixed values $r$%
, we introduce in (\ref{FCcylext}) a new integration variable $y=x/a$ and
expand the integrand in powers of $a$. The main contribution comes from the
mode $j=1/2$ and we have the leading term given below%
\begin{equation}
\langle \bar{\psi}\psi \rangle _{\mathrm{cyl}}\approx \frac{2^{1-q}q(a/r)^{q}%
}{\pi ^{2}\Gamma ^{2}((q+1)/2)r^{3}}\int_{mr}^{\infty
}dx\,x^{q}[(x^{2}-2m^{2}r^{2})K_{(q-1)/2}^{2}(x)+x^{2}K_{(q+1)/2}^{2}(x)].
\label{FCsmalla}
\end{equation}%
For a massless fermionic field, by using the result from \cite{Prud86} for
the integral involving the square of the MacDonald function, from here we
find%
\begin{equation}
\langle \bar{\psi}\psi \rangle _{\mathrm{cyl}}\approx \frac{q(q+1)}{2\pi
^{2}r^{3}}\left( \frac{a}{r}\right) ^{q},\;a/r\ll 1.  \label{FClargerm0}
\end{equation}

At large distances from the cylindrical boundary, for a massive field under
the condition $mr\gg 1$ the main contribution into the integral in (\ref%
{FCcylext}) comes from the lower limit of the integration and to the leading
order we find%
\begin{equation}
\langle \bar{\psi}\psi \rangle _{\mathrm{cyl}}\approx \frac{q\sqrt{mr}%
e^{-2mr}}{4\sqrt{\pi }r^{3}}\sum_{j}{\mathrm{Im}}\bigg[\frac{\bar{I}%
_{qj-1/2}(ma)}{\bar{K}_{qj-1/2}(ma)}\bigg].  \label{FClarger}
\end{equation}%
Here the imaginary part is easily taken by using Eq. (\ref{IbKb}). As we
could expect, in this limit the VEV is exponentially suppressed. At large
distances and for a massless field the behavior of the fermionic condensate
is described by Eq. (\ref{FClargerm0}). Note that the decreasing of the
fermionic condensate at large distances is stronger than in the case when
the string is absent. For points near the boundary, by using the uniform
asymptotic expansions for the modified Bessel functions, we can see that the
leading term in the asymptotic expansion of the fermionic condensate over
the distance from the boundary is given by the same expression (\ref{FCnear}%
) as in the interior region.

\subsection{VEV\ of the energy-momentum tensor}

The VEV for the energy-momentum tensor in the exterior region is found in
the way similar to that for the fermionic condensate. Here we omit the
details of the calculations and give the final result. The VEV is decomposed
into the sum of boundary-free and boundary-induced parts in the form given
by (\ref{EMT1}). In the region outside the cylindrical shell the
boundary-induced part is (no summation over $\nu $)%
\begin{equation}
\left\langle T_{\nu }^{\nu }\right\rangle _{\mathrm{cyl}}=\frac{q}{\pi
^{2}a^{4}}\sum_{j}\int_{\mu }^{\infty }dx\,x^{3}{\mathrm{Re}}\bigg[\frac{%
\bar{I}_{\beta _{j}}(x)}{\bar{K}_{\beta _{j}}(x)}F_{\beta _{j}}^{\mathrm{(ex)%
}(\nu )}(x,xr/a)\bigg].  \label{EMText}
\end{equation}%
In this formula we introduced the notations%
\begin{eqnarray}
F_{\beta _{j}}^{\mathrm{(ex)}(0)}(x,y) &=&\frac{\mu ^{2}/x^{2}-1}{2}%
\sum_{\delta =\pm 1}\delta \bigg(1+\frac{i\delta \mu }{\sqrt{x^{2}-\mu ^{2}}}%
\bigg)K_{qj-\delta /2}^{2}(xr/a),  \notag \\
F_{\beta _{j}}^{\mathrm{(ex)}(1)}(x,y) &=&K_{\beta _{j}}^{2}(y)-K_{\beta
_{j}+1}^{2}(y)+(2qj/y)K_{\beta _{j}}(y)K_{\beta _{j}+1}(y),  \label{Fbetext}
\\
F_{\beta _{j}}^{\mathrm{(ex)}(2)}(x,y) &=&-(2qj/y)K_{\beta _{j}}(y)K_{\beta
_{j}+1}(y),  \notag
\end{eqnarray}%
and $F_{\beta }^{\mathrm{(ex)}(3)}(x,y)=F_{\beta }^{\mathrm{(ex)}(0)}(x,y)$.
As an additional check we can see that these VEVs satisfy the trace relation
and the covariant conservation equation. By using formula (\ref{IbKb}), we
can write the vacuum densities in the form%
\begin{eqnarray}
\left\langle T_{0}^{0}\right\rangle _{\mathrm{cyl}} &=&\frac{q}{2\pi
^{2}a^{4}}\sum_{j}\int_{\mu }^{\infty }dx\,x(1-\mu ^{2}/x^{2})  \notag \\
&&\frac{W_{\beta _{j}}(x)[K_{\beta _{j}}^{2}(xr/a)-K_{\beta
_{j}+1}^{2}(xr/a)]+2\mu K_{\beta _{j}+1}^{2}(xr/a)}{K_{\beta
_{j}+1}^{2}(x)+K_{\beta _{j}}^{2}(x)-2(\mu /x)K_{\beta _{j}}(x)K_{\beta
_{j}+1}(x)},  \label{T00ext}
\end{eqnarray}%
for the energy density and in the form (no summation over $\nu $)

\begin{equation}
\left\langle T_{\nu }^{\nu }\right\rangle _{\mathrm{cyl}}=-\frac{q}{\pi
^{2}a^{4}}\sum_{j}\int_{\mu }^{\infty }dx\,\frac{x[W_{\beta _{j}}(x)-\mu
]F_{\beta _{j}}^{\mathrm{(ex)}(\nu )}(x,xr/a)}{K_{\beta
_{j}+1}^{2}(x)+K_{\beta _{j}}^{2}(x)-2(\mu /x)K_{\beta _{j}}(x)K_{\beta
_{j}+1}(x)},  \label{Tnuext}
\end{equation}%
for the radial and azimuthal stresses, $\nu =1,2$. We recall that the
summation over $j$ in these formulae goes in accordance with Eq. (\ref%
{SumjNot}).

In the case of a massless fermionic field from (\ref{EMText}) we find the
following expressions (no summation over $\nu $)%
\begin{eqnarray}
\left\langle T_{\nu }^{\nu }\right\rangle _{\mathrm{cyl}} &=&\frac{q}{\pi
^{2}a^{4}}\sum_{j}\int_{0}^{\infty }dx\,x^{3}F_{\beta _{j}}^{\mathrm{(ex)}%
(0,\nu )}(xr/a)  \notag \\
&&\times \frac{I_{\beta _{j}}(x)K_{\beta _{j}}(x)-I_{\beta
_{j}+1}(x)K_{\beta _{j}+1}(x)}{K_{\beta _{j}}^{2}(x)+K_{\beta _{j}+1}^{2}(x)}%
,  \label{EMTextm0}
\end{eqnarray}%
where%
\begin{equation}
F_{\beta _{j}}^{\mathrm{(ex)}(0,0)}(y)=\frac{1}{2}[K_{\beta
_{j}+1}^{2}(xr/a)-K_{\beta _{j}}^{2}(xr/a)],  \label{F0ext}
\end{equation}%
and for the corresponding functions for the radial and azimuthal stresses we
have $F_{\beta }^{\mathrm{(ex)}(0,\nu )}(y)=F_{\beta }^{\mathrm{(ex)}(\nu
)}(x,y)$, $\nu =1,2$.

Now we turn to the investigation of the VEV in the energy-momentum tensor
induced by the cylindrical shell in the exterior region in limiting cases.
First let us consider the limit $a\rightarrow 0$ for fixed values $r$.
Expanding the integrands in powers of $a$, we can see that the main
contribution comes from the terms with $j=1/2$, and the leading terms are
given by the expressions (no summation over $\nu $)%
\begin{eqnarray}
\left\langle T_{0}^{0}\right\rangle _{\mathrm{cyl}} &\approx &\frac{%
2^{1-q}qm(a/r)^{q}}{\pi ^{2}\Gamma ^{2}((q+1)/2)r^{3}}\int_{mr}^{\infty
}dx\,x^{q}(x^{2}-m^{2}r^{2})K_{(q-1)/2}^{2}(x),  \notag \\
\left\langle T_{\nu }^{\nu }\right\rangle _{\mathrm{cyl}} &\approx &-\frac{%
2^{1-q}qm(a/r)^{q}}{\pi ^{2}\Gamma ^{2}((q+1)/2)r^{3}}\int_{mr}^{\infty
}dx\,x^{q+2}F_{(q-1)/2}^{\mathrm{(ex)}(\nu )}(x,x),  \label{EMTsmalla}
\end{eqnarray}%
with $\nu =1,2$. For a massless field these terms vanish. In this case from
Eq. (\ref{EMTextm0}) we find the following leading behavior (no summation
over $\nu $):%
\begin{equation}
\left\langle T_{\nu }^{\nu }\right\rangle _{\mathrm{cyl}}\approx \frac{%
q^{2}(q+1)A_{\nu }}{\pi ^{2}(q-1)(q+2)r^{4}}\left( \frac{a}{r}\right) ^{q+1},
\label{EMTextLargm0}
\end{equation}%
with the coefficients%
\begin{equation}
A_{0}=\frac{q+3}{2(q+4)},\;A_{1}=\frac{1}{q+4},\;A_{2}=-1.  \label{A012}
\end{equation}%
For $q=1$ the VEVs behave like $(a/r)^{2}r^{-4}\ln (a/r)$.

At large distances from the cylinder and for a massive field the main
contribution comes from the lower limit of the integral in (\ref{EMText}).
By using the asymptotic formulae for the MacDonald function for large values
of the argument, we find%
\begin{eqnarray}
\left\langle T_{0}^{0}\right\rangle _{\mathrm{cyl}} &\approx &\frac{m}{2}%
\langle \bar{\psi}\psi \rangle _{\mathrm{cyl}},\;\left\langle
T_{1}^{1}\right\rangle _{\mathrm{cyl}}\approx -\frac{1}{2mr}\left\langle
T_{2}^{2}\right\rangle _{\mathrm{cyl}},  \notag \\
\left\langle T_{2}^{2}\right\rangle _{\mathrm{cyl}} &\approx &-\frac{%
qme^{-2mr}}{2\pi r^{3}}\sum_{j}qj{\mathrm{Re}}\bigg[\frac{\bar{I}%
_{qj-1/2}(ma)}{\bar{K}_{qj-1/2}(ma)}\bigg],  \label{EMTextLargr}
\end{eqnarray}%
for $mr\gg 1$. As we see, in this limit $\left\langle T_{1}^{1}\right\rangle
_{\mathrm{cyl}}\ll \left\langle T_{2}^{2}\right\rangle _{\mathrm{cyl}}\ll
\left\langle T_{0}^{0}\right\rangle _{\mathrm{cyl}}$. At large distances
from the cylinder and for a massless field the asymptotic behavior of the
boundary induced parts is given by formula (\ref{EMTextLargm0}).

The asymptotic behavior of the VEV\ for the energy-momentum tensor near the
cylindrical shell is found in a way similar to that for the interior region
and the leading terms are given by the formulae (\ref{T22NearCyl}), (\ref%
{T00nearCyl}). Hence, near the boundary the energy density and the azimuthal
stress in the interior and exterior regions have opposite signs, whereas the
radial stresses have the same sign. The boundary induced parts in the VEVs
of the energy density and the radial stress for the exterior region are
plotted in figures \ref{fig1}-\ref{fig3} as functions of the radial
coordinate and the parameters $q$ and $ma$.

\section{Conclusion}

\label{sec:Conc}

In this paper the vacuum polarization effects are investigated for a
fermionic field in the geometry of a cosmic string with a coaxial
cylindrical shell. We have assumed that on the shell the field obeys the MIT
bag boundary condition. In order to evaluate the fermionic condensate and
the VEV\ of the energy-momentum tensor one needs the complete set of
normalized eigenspinors satisfying the boundary condition. This set for the
region inside the cylindrical shell is considered in section \ref%
{sec:EigSpinor}. The corresponding mode-sums for both fermionic condensate
and the energy-momentum tensor contain series over the zeros of the
combination (\ref{ftilde}) of the Bessel function of the first kind and its
derivative. For the summation of these series we used a variant of the
generalized Abel-Plana formula previously derived in Ref. \cite{Saha04Mon}.
This formula allows us to extract from the respective VEVs the parts
corresponding to the cosmic string geometry without a cylindrical shell and
to present the part induced by the shell in terms of exponentially
convergent integrals for points away from the boundary. In this way the
renormalization procedure for the fermionic condensate and the
energy-momentum tensor is reduced to the renormalization of the
corresponding quantities in the geometry of the boundary-free cosmic string.
The renormalized VEV of the energy-momentum tensor for a fermionic field in
the boundary-free geometry is well investigated in literature. In appendix %
\ref{sec:Appendix}, by using the Abel-Plana summation formula, we give
alternative integral representations for both fermionic condensate and the
energy-momentum tensor in the case of a massive field.

In the region inside the shell, the parts in the VEVs induced by the
presence of the cylindrical boundary are given by formula (\ref{FCcyl}) for
the fermionic condensate and by Eq. (\ref{T00int}), (\ref{Tnuint}) for the
vacuum energy densities and stresses. These formulae are further simplified
for a massless fermionic field with the vacuum densities given by Eqs. (\ref%
{FCm0}) and (\ref{EMTm0}). For points near the cylindrical shell the
boundary induced parts in the VEVs dominate over the boundary-free parts and
diverge on the cylindrical shell. These type of divergences are well known
in quantum field theory with boundaries and are investigated for various
bulk and boundary geometries. In the problem under consideration, the
leading terms in the asymptotic expansions in powers of the distance from
the boundary are given by Eq. (\ref{FCnear}) for the fermionic condensate
and by Eqs. (\ref{T22NearCyl}) and (\ref{T00nearCyl}) for the components of
the energy-momentum tensor. These leading terms do not depend on the planar
angle deficit and are the same as for a cylindrical boundary in the
Minkowski bulk. The boundary induced parts in the VEVs vanish on the string
axis for $q>1$ and are non-zero in the case of a cylindrical boundary in the
Minkowski bulk. Since the boundary-free part diverges on the string axis,
for points near the string it dominates. For large values of the parameter $%
q $, which corresponds to a large planar angle deficit, the boundary-induced
VEVs are suppressed by the factor $(r/a)^{q}$. The boundary induced VEVs
have non-trivial dependence on the mass of the field and, as it is
illustrated by figure \ref{fig3}, for a massive field the polarization
effects can be stronger than for a massless one.

Fermionic vacuum densities in the region outside a cylindrical shell with
MIT bag boundary condition are investigated in section \ref{sec:Exterior}.
Subtracting from the mode-sums the parts corresponding to the geometry of a
string without boundaries and by making use of a complex rotation, we have
derived explicit expressions for the boundary induced VEVs. The
corresponding parts in the fermionic condensate and the energy-momentum
tensor are given by Eqs. (\ref{FCcylext}), (\ref{T00ext}) and (\ref{Tnuext}%
). When the cylinder radius goes to zero, for a fixed value of the radial
distance, the boundary induced part in the VEV\ of the energy-momentum
tensor vanishes as $a^{q}$ for a massive field and as $a^{q+1}$ for a
massless one. At large distances from the cylindrical shell this part is
exponentially suppressed for a massive field and decay as $r^{-4}(r/a)^{q+1}$
in the case of a massless field. Note that in the latter case the
boundary-free part behaves as $r^{-4}$ and it dominates at large distances.
For points near the cylindrical shell the leading terms in the asymptotic
expansions in powers of the distance from the boundary are given by the same
formulae as for the interior region. In this limit the total VEV is
dominated by the boundary induced part. In dependence of the mass, the
vacuum stresses can be either positive or negative, whereas the energy
density is positive. In the special case $q=1$, from the formulae derived in
the present paper we obtain the fermionic Casimir densities for a
cylindrical boundary in the Minkowski spacetime.

We have considered the idealized geometry of a cosmic string with zero
thickness. A realistic model for cosmic string has a non-trivial structure
on a length scale defined by the phase transition at which it is formed. As
it has been shown in Refs. \cite{Alle90,Alle92,Alle96}, the internal
structure of the string may have non-negligible effects even at large
distances. Here we note that when the cylindrical boundary is present, the
VEVs of the physical quantities in the exterior region are uniquely defined
by the boundary conditions and the bulk geometry. This means that if we
consider a non-trivial core model with finite thickness $b<a$ and with the
line element (\ref{ds21}) in the region $r>b$, the results in the region
outside the cylindrical shell will not be changed. As regards to the
interior region, the formulae given in this paper are the first stage of the
evaluation of the VEVs and other effects could be present in a realistic
cosmic string. Note that from the point of view of the physics in the
exterior region the cylindrical surface with MIT bag boundary condition can
be considered as a simple model of non-trivial string core.

\section*{Acknowledgments}

E.R.B.M. and V.B.B. thank Conselho Nacional de Desenvolvimento Cient\'{\i}%
fico e Tecnol\'{o}gico (CNPq), FAPESQ-PB/CNPq (PRONEX) and FAPES-ES/CNPq
(PRONEX) for partial financial support. A.A.S. was supported by the Armenian
Ministry of Education and Science Grant No. 119 and by Conselho Nacional de
Desenvolvimento Cient\'{\i}fico e Tecnol\'{o}gico (CNPq). A.S.T. was
supported in part by grant NFSAT-CRDF UC-06/07.

\appendix

\section{Vacuum densities in the geometry of a cosmic string without
boundaries}

\label{sec:Appendix}

In this appendix we consider the renormalized fermionic condensate and the
VEV of the energy-momentum tensor in the cosmic string geometry when the
cylindrical shell is absent. For a massless field the vacuum energy-momentum
tensor was found in \cite{Frol87,Dowk87b}. Fermionic propagators for a
massive field are considered in Refs. \cite{Line95,More95}. In the case of a
massive field, a representation of the VEVs for the energy-momentum tensor
in terms of contour integrals is given in \cite{Beze06}. Here an alternative
integral formulae are given by applying to the corresponding mode-sums the
Abel-Plana formula. We will do these calculations by using the mode-sum
formulae (\ref{FC0b}) and (\ref{EMTs1}) for the boundary-free VEVs.

First let us consider the fermionic condensate. The renormalization is done
by subtracting the corresponding quantity for the Minkowski background. The
latter is obtained from (\ref{FC0b}) putting $q=1$. Substituting in the
corresponding formulae%
\begin{equation*}
\frac{1}{\omega }=\frac{2}{\sqrt{\pi }}\int_{0}^{\infty }dse^{-\omega
^{2}s^{2}},
\end{equation*}%
integrating over $k$ and $\lambda $, and introducing a new integration
variable $y=r^{2}/2s^{2}$, we find the following representation of the
renormalized fermionic condensate:%
\begin{eqnarray}
\langle \bar{\psi}\psi \rangle _{\mathrm{s,ren}} &=&\langle 0|\bar{\psi}\psi
|0\rangle _{\mathrm{s}}-\langle 0|\bar{\psi}\psi |0\rangle _{\mathrm{M}}=-%
\frac{m}{2\pi ^{2}r^{2}}\int_{0}^{\infty }dye^{-m^{2}r^{2}/y-y}  \notag \\
&&\times \sum_{\delta =\pm 1}\sum_{j}\,\left[ qI_{qj-\delta
/2}(y)-I_{j-\delta /2}(y)\right] .  \label{FC01}
\end{eqnarray}%
Next, we apply to the series over $j$ the Abel-Plana summation formula in
the form (see, for example, \cite{Most97,Saha07})
\begin{equation}
\sum_{n=0}^{\infty }f(n+1/2)=\int_{0}^{\infty }dx\,f(x)-i\int_{0}^{\infty }dx%
\frac{f(ix)-f(-ix)}{e^{2\pi x}+1}.  \label{Abel}
\end{equation}%
It is easily seen that for the summand in (\ref{FC01}) the first integral on
the right-hand side of (\ref{Abel}) vanishes and, hence, the divergent parts
are explicitly cancelled. Introducing the MacDonald function, we arrive to
the following expression%
\begin{equation}
\langle \bar{\psi}\psi \rangle _{\mathrm{ren}}=\frac{2m}{\pi ^{3}r^{2}}%
\int_{0}^{\infty }dy\,e^{-m^{2}r^{2}/2y-y}\int_{0}^{\infty }dx\,g(q,x){%
\mathrm{Im}}[K_{ix+1/2}(y)].  \label{FCren}
\end{equation}%
where the notation%
\begin{equation}
g(q,x)=\cosh (\pi x)\left[ \frac{1}{e^{2\pi x/q}+1}-\frac{1}{e^{2\pi x}+1}%
\right] ,  \label{gqx}
\end{equation}%
is introduced.

In a similar way we can find the formula for the renormalized VEV of the
energy-momentum tensor. For the renormalized energy density and the
azimuthal stress we have the representations given below:%
\begin{eqnarray}
\left\langle T_{0}^{0}\right\rangle _{\mathrm{s,ren}} &=&\frac{r^{-4}}{2\pi
^{2}}\int_{0}^{\infty }dy\,ye^{-m^{2}r^{2}/2y-y}\sum_{\delta =\pm
1}\sum_{j}\,\left[ qI_{qj-\delta /2}(y)-I_{j-\delta /2}(y)\right] ,  \notag
\\
\left\langle T_{2}^{2}\right\rangle _{\mathrm{s,ren}} &=&\frac{r^{-4}}{\pi
^{2}}\int_{0}^{\infty }dy\,ye^{-m^{2}r^{2}/2y-y}\sum_{\delta =\pm 1}\delta
\,\sum_{j}j\left[ q^{2}I_{qj-\delta /2}(y)-I_{j-\delta /2}(y)\right] .
\label{T00T22Free}
\end{eqnarray}%
By making use of summation formula (\ref{Abel}) to the series over $j$, we
find the following formulae for the renormalized VEVs:%
\begin{eqnarray}
\langle T_{0}^{0}\rangle _{\mathrm{s,ren}} &=&-\frac{2}{\pi ^{3}r^{4}}%
\int_{0}^{\infty }dy\,\left( y+m^{2}r^{2}\right)
e^{-m^{2}r^{2}/2y-y}\int_{0}^{\infty }dx\,g(q,x){\mathrm{Im}}[K_{ix+1/2}(y)],
\notag \\
\left\langle T_{2}^{2}\right\rangle _{\mathrm{s,ren}} &=&\frac{4}{\pi
^{3}r^{4}}\int_{0}^{\infty }dy\,ye^{-m^{2}r^{2}/2y-y}\int_{0}^{\infty
}dxxg(q,x){\mathrm{Re}}\left[ K_{ix+1/2}(y)\right] .  \label{T00sren}
\end{eqnarray}%
The radial stress is found from (\ref{FCren}) and (\ref{T00sren}) by using
the trace relation.

Formulae (\ref{T00sren}) are further simplified for a massless fermionic
field. The integration over $y$ is done by using the formula%
\begin{equation}
\int_{0}^{\infty }dy\,ye^{-y}K_{ix+1/2}(y)=\frac{\pi (4x^{2}+1)}{24\cosh
(\pi x)}(2ix+3).  \label{KInt}
\end{equation}%
Substituting this into Eq. (\ref{T00sren}) and integrating over $x$, we find
\begin{equation}
\left\langle T_{0}^{0}\right\rangle _{\mathrm{s,ren}}=-\frac{1}{3}%
\left\langle T_{2}^{2}\right\rangle _{\mathrm{s,ren}}=-\frac{%
(q^{2}-1)(7q^{2}+17)}{2880\pi ^{2}r^{4}}.  \label{T00srenm0}
\end{equation}%
In the massless case the radial stress is equal to the energy density.

\end{document}